# Observation of an electronic microemulsion phase emerging from a quantum crystal-to-liquid transition

Jiho Sung[1,2]†, Jue Wang[1,2]†, Ilya Esterlis[2,3]†, Pavel A. Volkov[2,4]†, Giovanni Scuri[1,2,5], You Zhou[6], Elise Brutschea[1], Takashi Taniguchi[7], Kenji Watanabe[7], Yubo Yang[8], Miguel A. Morales[8], Shiwei Zhang[8], Andrew J. Millis[8,9], Mikhail D. Lukin[2], Philip Kim[2,10], Eugene Demler[11]* & Hongkun Park[1,2]*

[1]Department of Chemistry and Chemical Biology, Harvard University; Cambridge, MA 02138, USA.
[2]Department of Physics, Harvard University; Cambridge, MA 02138, USA.
[3]Department of Physics, University of Wisconsin-Madison; Madison, WI 53706, USA.
[4]Department of Physics, University of Connecticut; Storrs, CT 06269, USA.
[5]E. L. Ginzton Laboratory, Stanford University; Stanford, CA 94305, USA.
[6]Department of Materials Science and Engineering, University of Maryland; College Park, MD 20742, USA.
[7]Research Center for Functional Materials, National Institute for Materials Science; 1-1 Namiki, Tsukuba 305-0044, Japan.
[8]Center for Computational Quantum Physics, Flatiron Institute; New York, NY 10010, USA.
[9]Department of Physics, Columbia University; New York, NY 10027, USA.
[10]John A. Paulson School of Engineering and Applied Sciences, Harvard University; Cambridge, MA 02138, USA.
[11]Institute for Theoretical Physics, ETH Zürich; CH-8093, Zürich, Switzerland.
†These authors contributed equally to this work
*Corresponding author. Email: hpark@g.harvard.edu; demlere@phys.ethz.ch

**Abstract:**

Strongly interacting electronic systems possess rich phase diagrams resulting from the competition between different quantum ground states. A general mechanism that relieves this frustration is the emergence of microemulsion phases, where regions of different phase self-organize across multiple length scales. The experimental characterization of these phases often poses significant challenges, as the long-range Coulomb interaction microscopically mingles the competing states. Here, we use cryogenic reflectance and magneto-optical spectroscopy to observe the signatures of the mixed state between an electronic Wigner crystal and an electron liquid in a $MoSe_2$ monolayer. We find that the transit into this "microemulsion" state is marked by anomalies in exciton reflectance, spin susceptibility, and Umklapp scattering, establishing it as a distinct phase of electronic matter. Our study of the two-dimensional electronic microemulsion phase elucidates the physics of novel correlated electron states with strong Coulomb interactions.

The interplay between Coulomb interactions and kinetic energy is at the heart of correlated electron physics and underlies the emergence of many exotic phases of matter. Despite a plethora of complex phenomena, such systems share general principles. Of particular importance is the fact that long-range Coulomb forces forbid direct first-order phase transitions, which are instead replaced by intermediate phases with intricate meso- or nano-scale structures (*1-7*). Such ideas have been proposed to explain the phase diagrams of strongly correlated materials, including the high transition temperature superconductors (*8*), colossal magnetoresistance manganates (*9, 10*), and excitonic Wigner crystal/superfluid phase in semiconductors (*11, 12*). Similar physics may be relevant for other quantum systems with long-range interactions, such as ultracold polar molecules with dipolar interactions (*13, 14*) and magnetic atoms (*15*). However, a direct confirmation of electronic mixed phases is lacking in solid-state systems because crystalline lattice transformations often coincide with electronic phase transitions (*9, 10*). Theoretical and experimental characterization of mixture phases remains challenging due to the multi-scale nature of the electronic order.

Here we focus on the case of the density-driven crystal-to-liquid transition in a low-density two-dimensional electron gas (2DEG) hosted in an atomically thin semiconductor. The low electron densities in these systems ensure that electronic transitions do not trigger lattice instabilities, making it a model correlated electron system with negligible lattice effects. In the low-density regime, Quantum Monte Carlo studies predict that electrons spontaneously arrange into a crystalline solid – the Wigner crystal – when the ratio of the Coulomb interaction to the kinetic energy, $r_s$ $(= m_e^* e^2/(4\pi\varepsilon_0 \varepsilon \hbar^2 \sqrt{\pi n}))$ is around 30 (*16, 17*). Here, $m_e^*, e, \varepsilon_0, \varepsilon, \hbar$ and $n$ denote the effective mass of electrons, elementary charge, vacuum permittivity, dielectric constant, reduced Planck constant, and electron density. With increasing electron density, this Wigner crystal melts into a liquid due to increasing quantum fluctuations.

Landau-Ginzburg theory that neglects the long ranged Coulomb interaction suggests that the crystal-to-liquid transition is of first order and proceeds via macroscopic phase separation (*18*). However, such macroscopic density inhomogeneity is forbidden in an electron system with unscreened long-range interactions due to the associated large Coulomb energy penalty. Instead, it is predicted that quantum melting of an electron Wigner crystal proceeds via a sequence of intermediate microemulsion phases, *e.g.*, bubbles of one phase embedded in another phase or alternatively stripe-like liquid crystalline phases (Fig. 1A), with associated continuous phase transitions (*1-3, 7*). Other possibilities for intermediate phases, such as metallic density wave states, have also been proposed (*19, 20*). The magnetic structure in this intermediate phase can also be intricate due to the highly frustrated nature of magnetic interaction (*21, 22*).

Previous studies of the transition from the liquid phase at higher densities to the crystalline phase at lower densities primarily relied on transport measurements (*23-28*). Recently, transport signatures of unconventional magnetic behavior have been reported close to the metal-insulator transition (*28*). However, such measurements do not allow the identification of intermediate microemulsion phases because the signatures of microemulsion phases in transport are extremely challenging to predict theoretically (*29*). Moreover, thermodynamic probes that can identify different electronic phases and transitions between them can be more desirable to disentangle the electron correlation effect from disorder-induced effect.

**Electron Wigner crystal in a MoSe$_2$ monolayer**

The recent discovery of Wigner crystal phases in atomically thin transition metal dichalcogenides offers a new avenue to investigate the fundamental questions regarding new phases of matter near the quantum crystal-to-liquid transition (*30, 31*). In these materials, optically generated excitons are sensitive to both charge and spin order of the surrounding electrons (*30-33*), thus providing a new local detection scheme that provides insights into the electronic phase diagram.

In our experiment, we use dilution refrigerator-based scanning confocal microscopy to probe the melting of an electron Wigner crystal formed in a MoSe$_2$ monolayer encapsulated by hexagonal boron nitride (hBN) (Fig. 1, A and B). We conduct circular polarization resolved spectroscopy measurements as a function of electron density (tuned by a gate voltage), perpendicular magnetic field, and temperature (Methods). In a MoSe$_2$ monolayer, the lowest energy optical transitions occur in the K (K′) valleys. In these transitions, σ+ (σ−) polarized light selectively couples to up (down) spin electrons owing to the optical selection rules and spin-valley locking (Fig. 1C) (*33*). The exciton-electron interaction depends on whether a K (K′) valley exciton interacts with either a K' (K) valley electron (case 1) or a K (K′) valley electron (case 2). In case 1, the interaction is strong, leading to the emergence of higher energy repulsive polaron (RP) and lower energy attractive polaron (AP) branches in the spectrum (*34*). In case 2, the interaction is comparatively weaker and is a combination of exchange interaction and Pauli blocking (*33, 34*). The difference between intra- and inter-valley exciton-electron interaction enables the detection of electron spin polarization by circularly polarized light.

Figure 1E and F show representative reflectance contrast spectra $R_C$ (defined in Supplementary Text) at $B = 9$ T with left ($\sigma^-$) and right ($\sigma^+$) circularly polarized light as a function of electron density. A perpendicular magnetic field lifts the degeneracy of the band minima, so the electrons doping the MoSe$_2$ monolayer are spin and valley polarized at low electron doping (in this case, with spin down in the K' valley as shown in Fig. 1C). The $\sigma^+$-reflectance contrast spectrum that probes the K valley (Fig. 1F) shows two polaron branches due to inter-valley exciton-electron interaction, while the $\sigma^-$ spectrum probing the K' valley does not (Fig. 1E). At higher doping densities ($n > 2.35 \times 10^{12}$ cm$^{-2}$), the electron starts to fill both valleys, and two polaron branches are present in both spectra. The inequivalence between $\sigma^+$ and $\sigma^-$ spectra reveals that the system is still partially spin-polarized.

The density dependence of the exciton reflectance reveals two important features: The optical response is not a smooth function of the electron density. Instead, it is segmented by anomalies, the most prominent of which occurs near $n_* = 0.82 \times 10^{12}$ cm$^{-2}$ (upward arrows in Fig. 1, H and I). Because the excitonic properties are sensitive to the charge and spin properties of the surrounding electrons, these anomalies are indications of a dramatic change in the electronic state. The detection of this previously unobserved feature is facilitated by the device being mounted inside a dilution refrigerator with a base lattice temperature of 16 mK (the electron temperature is higher due to light absorption, as discussed below) and an excitation light power below 0.7 nW (see Materials and Methods).

At low electron densities, we observe a higher-energy spectral feature (Fig. 1G). At electron densities below $n_{WC} = 0.35 \times 10^{12}$ cm$^{-2}$, this feature blueshifts linearly with electron density (along the black dashed line in Fig. 1G) as seen previously (*30*). This higher-energy resonance originates from the umklapp scattering of excitons by the periodic potential created by the electron Wigner crystal: the exciton state with momentum $k = k_{WC}$ ($k_{WC}$ denotes the reciprocal lattice vector of the electron Wigner crystal) is folded back to the zero-momentum light cone and acquires a finite oscillator strength (Fig. 1D) (*30*). Notably, at densities greater than $n_{WC}$ the feature changes slope and extends to significantly higher densities than previously observed, albeit with a significant broadening. At sufficiently high electron densities (Fig. 1, G and H), the feature vanishes.

Taken together, the spectral anomalies and the extended umklapp feature demonstrate an unexpected evolution of the electronic state between a simple Wigner crystal and liquid phases. Before investigating the behavior in this intermediate density range, we first establish that the magnetic properties of the system at sufficiently high and low electron densities are those of a Fermi liquid and Wigner crystal, respectively.

**Fermi liquid and Wigner crystal**

The magnetic response of a Fermi liquid can be characterized by the critical density for full spin polarization at non-zero magnetic fields and the spin susceptibility near zero field. As discussed in the preceding section, the filling of electrons into the K valley results in the inter-valley polaronic dressing of K′-valley excitons, and vice versa. We can therefore use the onset of the $\sigma^-$ AP resonance in the reflectance spectra (Fig. 1, E and F) to determine the critical density at which the fully spin-polarized liquid starts to become partially polarized (see fig. S1 to S3) (*33*).

Figure 2A shows the critical densities determined by the onset of the $\sigma^-$ AP resonance at various magnetic fields. For a given magnetic field, we find the system is fully spin-polarized below a critical electron density, as expected for a Fermi liquid (35). The critical densities extracted from our experiments are in good agreement with predictions from a fixed-node diffusion quantum Monte Carlo (QMC) model (*17, 35*) for a clean 2DEG (Fig. 2A). We note that the critical density we observe is an order of magnitude higher than that predicted for non-interacting electrons in MoSe$_2$ (*33*), demonstrating the strong Coulomb interactions present in our system (fig. S4).

To extract the spin susceptibility of the 2DEG at low magnetic fields we obtain the magneto-optical signal $\widetilde{M} = (I^{\sigma+}/I^{\sigma+}_{max} - I^{\sigma-}/I^{\sigma-}_{max})/(I^{\sigma+}/I^{\sigma+}_{max} + I^{\sigma-}/I^{\sigma-}_{max})$, which is the difference in normalized reflected intensity between $\sigma^{\pm}$ light divided by their sum. Here, $I$ ($I_{max}$) denotes the reflected intensity in the MoSe$_2$ monolayer region (at a density of $1.0 \times 10^{13}$ cm$^{-2}$). We compute $\widetilde{M}$ using reflectance spectra in a narrow energy range, 5 meV red-detuned from the main exciton resonance at zero doping. Optical selection rules imply that $\widetilde{M}$ is proportional to the imbalance of electrons in the lowest conduction bands (Fig. 1C), which, owing to spin-valley locking, can be interpreted as the spin (or valley) polarization of the electrons (see fig. S5 and S6 for details) (*32, 36*). The field dependence of $\widetilde{M}(H)$ further allows us to extract the spin susceptibility $\chi$ of electrons shown in Fig. 2B, which is normalized to the spin susceptibility of the non-interacting 2DEG, $\chi_0 = \left(\frac{g\mu_B}{2}\right)^2 \frac{m^*}{\pi\hbar^2}$ (*37*). (See fig. S7 and S8 for other methods to obtain the spin susceptibility, which yield similar results). As anticipated, $\chi/\chi_0$ grows with decreasing electron density because the Coulomb interaction favours spin polarization.

Similar to the critical field for full polarization, this experimentally extracted spin susceptibility agrees well with predictions from QMC studies (*17, 35*) as shown in the dashed line in Fig. 2B, further confirming that the high density electron system is well-described as a clean, strongly interacting 2D Fermi liquid. A Fermi liquid should furthermore exhibit a temperature-independent spin susceptibility (*38*) for $T \ll E_F$, where $E_F$ is the (renormalized) Fermi energy. As will be described below, we indeed observe a temperature-independent susceptibility above a certain density denoted by $n_{\mathrm{FL}}$ (Fig. 3D).

The magnetic response of the low-density ($n < n_{\mathrm{WC}}$) Wigner crystal is drastically different from that of a Fermi liquid. In the Wigner crystal phase, the electrons are localized in real space so that the spins are correlated only by the (extremely weak) exchange interaction (*21, 39*). For temperatures above the exchange interaction scale but below the melting temperature of the Wigner crystal, the magnetic behavior should thus be that of independent spins, with magnetization following the Brillouin function $M = \frac{1}{2}g\mu_B n \tanh\left(\frac{g\mu_B H}{2k_B T_{elec}}\right)$ and an associated Curie susceptibility $\chi = \left(\frac{g\mu_B}{2}\right)^2 \frac{n}{k_B T}$ (*2, 7*). In sharp contrast to the liquid phase, the susceptibility is expected to increase with electron density and to depend strongly on temperature.

To probe the magnetism of the low-density regime, where light-indued heating becomes important, we use a continuous-wave laser at a single energy (1.636 eV, white dashed lines in Fig. 1, E and F) below the exciton resonance to obtain $\widetilde{M}$ as a function of electron density and magnetic field (Materials and Methods): this approach reduces the light power reaching the sample. By fitting the measured $\widetilde{M}(n, H)$ at different temperatures using the Brillouin function, we determine the conduction band *g*-factor and the electron temperature ($T_{\mathrm{elec}}$). With a 60 fW diffraction-limited laser spot, we reach an electronic temperature of 80 mK with the device at a base lattice temperature of 16 mK (fig. S9). When the temperature exceeds 150 mK, the extracted electron temperatures approach the lattice temperatures (fig. S10).

Figure 2C shows a plot of $\widetilde{M}$ normalized by the electron density ($\widetilde{M}/n$) as a function of magnetic field normalized by the electron temperature ($\mu_B H/k_B T_{elec}$) in the Wigner crystal regime ($n < n_{WC} = 0.35 \times 10^{12}$ cm$^{-2}$). As clearly seen in the figure, the data at different electron densities and electron temperatures collapse onto a single curve. When we extract the spin susceptibility at various temperatures from linear fits to $\widetilde{M}(H)$ at small fields, we find an inverse relationship between the spin susceptibility and electron temperature (Fig. 2D). The successful application of the scaling analysis across different electron densities and temperatures demonstrates that the behavior of the electrons in the low-density regime ($n < n_{WC}$) is well-described by a Wigner crystal with an independent spin localized at each lattice site.

**Microemulsion phase characterized by crystal-liquid coexistence**

Following the demonstration of a conventional Fermi liquid at high density and a Wigner crystal at low density, we proceed to characterize charge and spin properties of the intermediate density range in our MoSe$_2$ system. We begin with a detailed analysis of the exciton umklapp scattering, which we introduced in the previous section (Fig. 1, D, G and H). To further enhance the relevant features, we subtract the fitted main exciton spectral profile from the reflectance contrast, take the derivative with respect to the electron density, and plot it against the energy detuning from main exciton peak (Fig. 3A). We then determine the energy splitting between the higher energy umklapp feature and the main exciton peak as well as the umklapp linewidth, as shown in Fig. 3, B and C (See also fig. S11 and S12).

When $n_{WC} < 0.35 \times 10^{12}$ cm$^{-2}$, the umklapp linewidth is close to the main exciton linewidth, while the energy splitting shows a linear increase with the electron density $\Delta E_u = \frac{h^2 n}{\sqrt{3} m_X}$, confirming that all electrons are crystalized into a triangular lattice Wigner crystal (*30*). However, when $n > n_{WC}$, $\Delta E_u$ becomes only weakly dependent on density, suggesting that the periodicity of the underlying electron lattice is approximately independent of $n$. As the electron density approaches $n_*$, the density at which the exciton intensity anomaly is observed in Fig. 1, H and I (see also fig.

S13), the width of the umklapp feature increases rapidly (bottom panel of Fig. 3C). The width of the umklapp scattering arises from the finite spatial extent of crystalline correlations: this length scale can be estimated as $l_{\text{corr}}/a_{\text{WC}} \sim \Delta E_{\text{u}}/\text{linewidth}$, where $l_{\text{corr}}$ is the correlation length, representing the radius of the domain, and $a_{\text{WC}}$ is the lattice constant of the Wigner crystal. We obtain $l_{\text{corr}} \sim 3a_{\text{WC}}$ in the low-density regime ($n < n_{\text{WC}}$), indicating the presence of well-defined crystalline regions as in the previous report (30). Between $n_{\text{WC}}$ and $n_*$, $l_{\text{corr}}$ decreases gradually and drops below $a_{\text{WC}}$ at around $n_*$, illustrating the continuous weakening of the crystalline correlations or a decrease in the size of the Wigner crystal domains.

In addition to the spectral anomalies and the umklapp feature, which provide insight into the evolution of the charge degree of freedom, we also extract spin susceptibility from linear fits to $\widetilde{M}(H)$ curves measured using the single energy (1.636 eV) excitation, as described in the preceding section. In Fig. 3D, we show the evolution of the reduced spin susceptibility ($\chi/\chi_0$) as a function of electron density. We observe two abrupt changes in the slope of the spin susceptibilities near $n_{\text{WC}}$ and $n_*$: below $n_{\text{WC}}$, the spin susceptibility follows a Curie law $\chi \propto \frac{n}{T}$, as expected for independent spins localized to Wigner crystal lattice sites, while above $n_*$ the susceptibility decreases monotonically with density for all temperatures, consistent with a Fermi-liquid-like response (Fig. 3E). From this susceptibility data we infer the temperature evolution of the critical densities, $n_{\text{WC}}(T)$ and $n_*(T)$. Between $n_{\text{WC}}(T)$ and $n_*(T)$, the spin susceptibility shows an essentially linear dependence on density, but with a different slope than the Curie susceptibility (see fig. S14).

The data in Figs. 2 and 3 suggest two phase boundaries at $n_{\text{WC}}(T)$ and $n_*(T)$ that separate three phases: a Wigner crystal phase for $n < n_{\text{WC}}$, a liquid phase for $n > n_*$, and the intermediate phase occupying the range $n_{\text{WC}} < n < n_*$. The behavior in the intermediate phase is consistent with a state in which only a fraction of the electrons participate in the formation of the crystal, with a lattice constant associated to the density $n_{\text{WC}}$. The persistence of exciton umklapp scattering implies the size of the crystalline regions remains appreciable. The spin susceptibility in the intermediate phase is smaller than that expected for a Wigner crystal but, at low temperatures, is significantly enhanced relative to the high-density liquid. To first order, the susceptibility is well-described by a linear interpolation between its values at $n_{\text{WC}}$ and $n_*$ (Fig. 3, D and E). Taken together with the umklapp evolution, these observations indicate that the intermediate phase is a microemulsion phase with nano- or meso-scale mixture of liquid and crystal regions, with the areal fractions of the two phases evolving with density according to the lever rule of two-phase coexistence (*1-3, 7*). It is important to note that the intermediate phase is flanked by pronounced anomalies in a number of spectral observables, suggesting that it is a distinct thermodynamic phase rather than a broad crossover between the crystal and liquid.

**Phase diagram and Pomeranchuk effect**

The full temperature and density dependence of the data is summarized in the ($n$, $T$) phase diagram displayed in Fig. 4 (see fig. S15 for estimating electron temperature associated with $n_{\mathrm{WC}}$ and $n_*$ from the reflectance measurement). The phase diagram includes a conventional Wigner crystal for $n < n_{\mathrm{WC}}(T)$, the new intermediate phase(s) for $n_{\mathrm{WC}}(T) < n < n_*(T)$, and a liquid for $n > n_*(T)$. The liquid region may be further divided into two distinct regions: Figure 3D shows that while the spin susceptibility is essentially temperature-independent at the highest electron densities, it acquires a strong temperature dependence (Fig. 3D inset) below a characteristic density near $1.5 \times 10^{12}$ cm$^{-2}$. We define the characteristic density $n_{\mathrm{FL}}(T)$ as that at which the spin susceptibility deviates from the highest temperature (1.64 K) value. For densities $n_*(T) < n < n_{\mathrm{FL}}(T)$, the spin susceptibility still increases with decreasing electron density but is in general larger than that expected for a conventional Fermi liquid. We note that near $n_{\mathrm{FL}}$ the exciton features evolve smoothly and there are no pronounced anomalies in the spin susceptibility.

At temperatures below 1 K, both boundaries of the intermediate phase exhibit a rightward slant in the *n-T* phase diagram, indicative of the Pomeranchuk effect associated with enhanced stability of the crystalline phase upon heating due to its large spin entropy (*1, 7, 40-43*). The overall magnitude of the effect is consistent with expectations based on QMC calculations (fig. S16) (*35*). The widening of the coexistence region upon increasing temperature also agrees qualitatively with theoretical predictions, although the width of the coexistence region is strongly underestimated.

**Discussion and outlook**

Our experiments demonstrate that the quantum melting of the Wigner crystal in a MoSe$_2$ monolayer proceeds via an intermediate microemulsion phase characterized by nano- or meso-scale coexistence of electron crystal and liquid. The unexpectedly large range of densities where the intermediate phase is stable suggests that it is as an important ingredient in the physics of strongly correlated electron systems and may have implications in other puzzles involving 2DEGs (*25, 26*).

Previous theoretical studies (*1-3, 7*) have shown that, because of the long-range Coulomb interactions, macroscopic phase separation is forbidden in Wigner crystal melting, leading to intricate mesoscale structures. The lever-rule mixing we observe between $n_{\text{WC}}$ and $n_*$ suggests that the system behavior is well described by mesoscale domains of the liquid and crystal phases, although the detailed organization of these domains remains to be characterized in future work. Near $n_{\text{WC}}$ and $n_*$, new phases were predicted in which the minority component appears in the form of "bubbles" of a fixed size (*1-3*). For sufficiently small bubble sizes, the system properties should be strongly affected by the associated interfaces, implying deviations from the macroscopic lever rule. Our data is consistent with this scenario, as the clear deviations from the lever rule are observed near $n_{\text{WC}}$ and $n_*$ (Fig. 3, D and E, see also fig. S14). While disorder may play a role in these observations, the relative sharpness of the phase boundaries indicates that the observed phases are likely not entirely due to strong disorder or inhomogeneity. It has been demonstrated that weak disorder enhances the intrinsic tendency toward crystallization (*44*) and may thus contribute to the widening of the coexistence region and persistent crystalline correlations we observe down to $r_s \sim 20$ (Fig. 4).

Our results also reveal unconventional behaviour of the 2DEG in the liquid phase. In the density range $n_*(T) < n < n_{\text{FL}}(T)$, the spin susceptibility (Fig. 3D) has much a stronger temperature and density dependence than QMC (*35*) and Fermi-liquid-based (*45*) predictions. In the same density range, the reflectance spectra also show some residual umklapp scattering with a linewidth that is significantly larger than in the proximate Wigner crystal and intermediate phases (Fig. 3, A to C). Such behavior may be due to local moments associated with residual crystallites or the more tantalizing possibility of a non-Fermi liquid regime driven by crystalline fluctuations near the onset of the inhomogeneous intermediate phase (*46*).

Our study serves as a starting point for investigating many more multi-scale ordered phases of electronic matter, such as microemulsions of magnetic, superconducting and charge ordered states (*47-49*). The properties of such phases have been scarcely studied and may harbor new functionalities, especially due to the prominent role of exotic inter-phase interfaces. The variety of correlated electronic phases observed recently in two-dimensional materials (*32, 50-55*) provides a natural platform for further exploration of microemulsion phases by exploiting their facile tunability via, e.g., multilayer heterostructures (*31*), or sample-gate distance (*56, 57*). Furthermore, local probes such as scanning tunneling microscopy (*51*) or a scanning electron transistor (*58*) will enable the characterization and control of microemulsion phases at the nanoscale.

**Acknowledgments**

We acknowledge useful discussions with S. Kivelson, A. Imamoglu, T. Smolenski, N. Leisgang, and A. A. Zibrov. We thank T. Smolenski for sharing some of his unpublished data with us.

**Funding:** We acknowledge support from AFOSR (FA9550-21-1-0216), the DoD Vannevar Bush Faculty Fellowship (N00014-16-1-2825 for H.P., N00014-18-1-2877 for P.K.), NSF CUA (PHY-1125846 for H.P., E.D., and M.D.L.), Samsung Electronics (for H.P. and P.K.), NSF (PHY-1506284 for H.P. and M.D.L., DGE-1745303 for E.B.), AFOSR MURI (FA9550-17-1-0002), ARL (W911NF1520067 for H.P. and M.D.L.), DOE (DE-SC0020115 for H.P. and M.D.L. and DE-SC0022885 for Y.Z.). E.D. acknowledges support by the SNSF Project No. 200021_212899. K.W. and T.T. acknowledge support from the JSPS KAKENHI (Grant Numbers 19H05790, 20H00354 and 21H05233). The Flatiron Institute is a division of the Simons Foundation.

**Author contributions:** H.P. and E.D. conceived the project. J.S., J.W., G.S., Y.Z. and E.B. fabricated the samples, and designed and performed the experiments. I.E. and P.A.V. performed calculations. I.E., P.A.V., Y.Y., M.A.M., S.Z., A.J.M. and E.D. contributed to theoretical descriptions. J.S., J.W., I.E., and P.A.V analysed the data. T.T. and K.W. provided hBN samples. J.S., J.W., I.E., P.A.V., E.D. and H.P. wrote the manuscript with extensive input from the other authors. H.P., E.D., P.K. and M.D.L. supervised the project.

**Competing interests:** The authors declare no competing interests.

**Data and materials availability:** The data that support the plots within this paper and other findings of this study are available from the corresponding authors upon reasonable request.

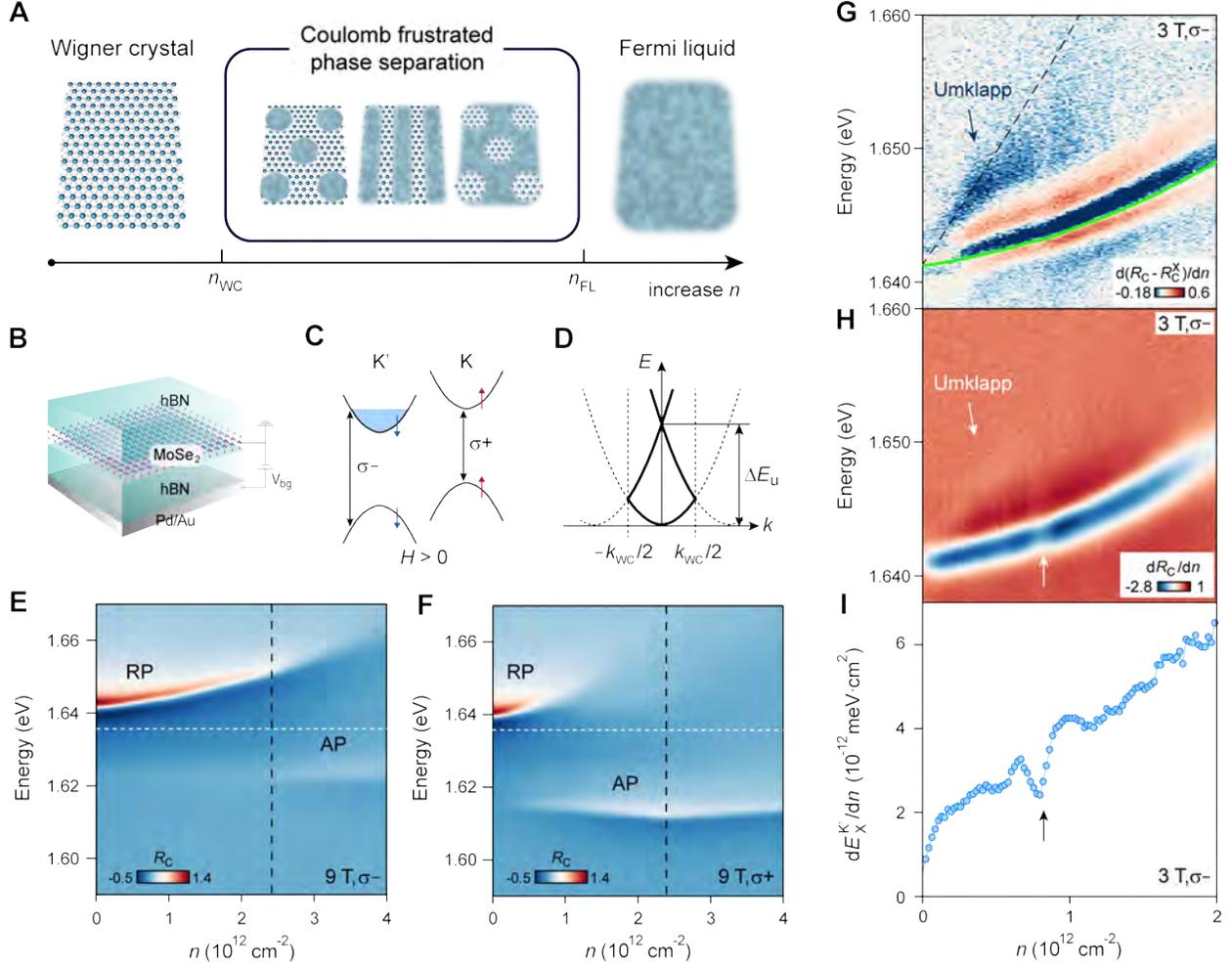

**Fig. 1. 2D electron phases and exciton spectroscopy.** (**A**) Schematic phase diagram of a 2D electron system including a Wigner crystal, a Fermi liquid, and the intermediate phases composed of meso- or nano- scale regions containing the two competing phases resulting from Coulomb frustrated phase separation. (**B**) Schematic of the device structure. A MoSe$_2$ monolayer is encapsulated by hBN and grounded. A bottom gate voltage, $V_{bg}$ is applied to the Pd/Au back gate to control the electron density in a MoSe$_2$ monolayer. (**C**) Schematic showing the relative energy alignment of the electronic band extrema at the K and K' valleys under a positive magnetic field. It also shows the spin-valley locking and the valley optical section rule. (**D**) Schematic of the exciton dispersion showing the higher-energy resonance at exciton momentum $k = 0$, arising from umklapp scattering by the periodic potential of the electron Wigner crystal. The energy splitting $\Delta E_u$ from the main exciton resonance is determined by reciprocal Wigner crystal lattice vector $k_{WC}$. (**E** and **F**) Left ($\sigma^-$) (E) and right ($\sigma^+$) (F) circularly polarized reflectance contrast spectra at 9 T and at a base lattice temperature of 16 mK as a function of electron density under light power of 0.7 nW. Black dashed lines show the density ($2.35\times10^{12}$ cm$^{-2}$) above which electrons start to fill the second valley (K valley) of opposite spin. White dashed lines mark the energy (1.636 eV) for obtaining the magneto-optical signal, $\widetilde{M}$. (**G**) Color map of the derivative of reflectance contrast spectra (3 T, $\sigma^-$) with respect to electron density. A Lorentz fitted exciton/RP peak was subtracted to emphasize the higher-energy features from umklapp scattering. The green line represents the fitted exciton/RP resonance energy and the black dashed line indicates the expected resonance energy from umklapp scattering of excitons. The difference between the two resonances correspond to the energy splitting $\Delta E_\mathrm{u}$. (**H**) Color map of the derivative of reflectance contrast spectra (3 T, $\sigma^-$) with respect to electron density, without subtracting the fitted exciton/RP Lorentzian. Note the clear anomaly near $0.82\times10^{12}$ cm$^{-2}$ and the higher energy umklapp feature. (**I**) Derivative of fitted exciton/RP resonance energy with respect to electron density shows a discontinuity around $0.82\times10^{12}$ cm$^{-2}$. Reflectance contrast spectra measurements in panels (G to I) were performed under light power of 70 pW at a base lattice temperature of 16 mK.

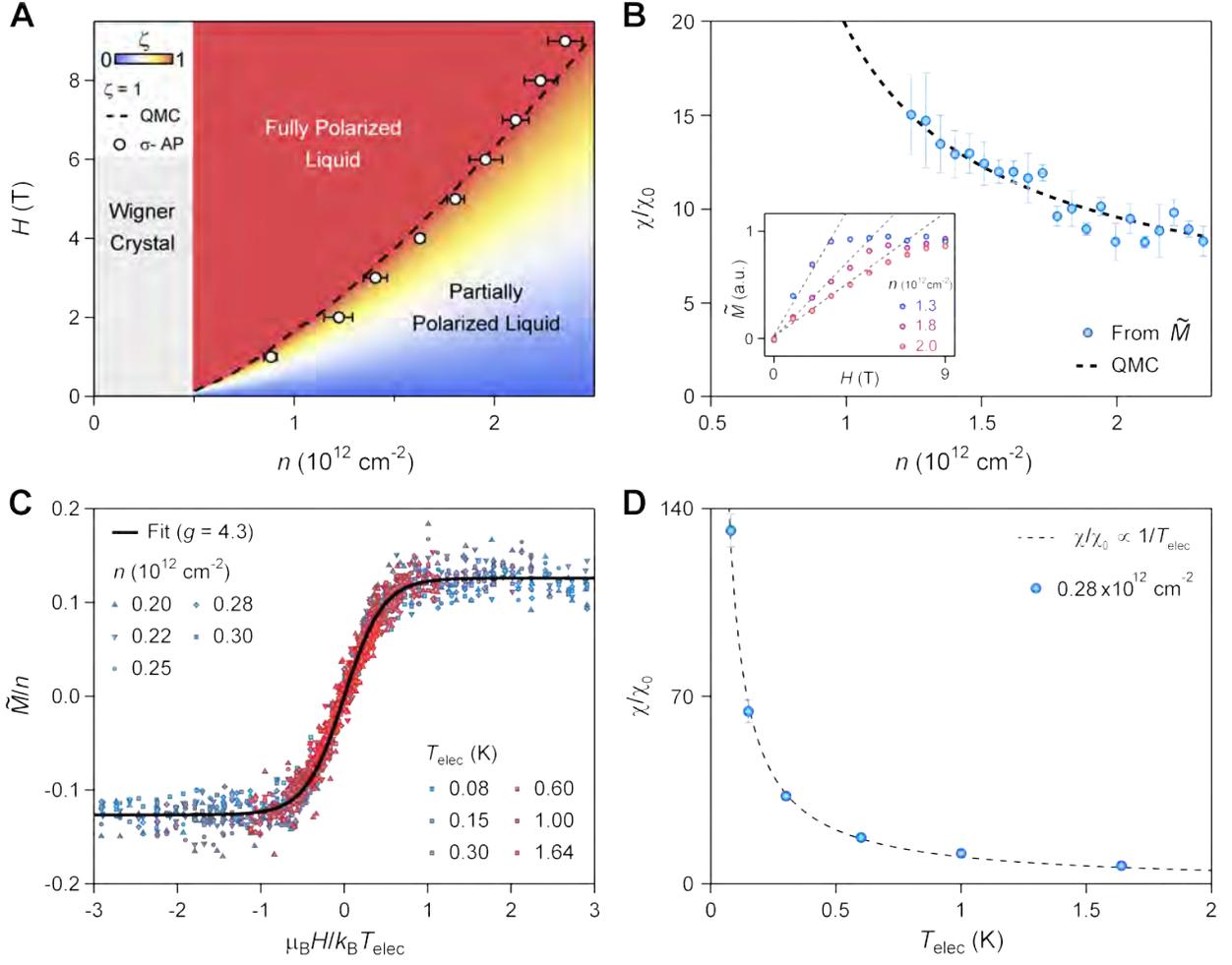

**Fig. 2. Fermi-liquid and Wigner-crystal magnetism.** (**A**) Spin polarization of the 2D electron system with respect to electron density and magnetic field. The color scale represents the spin polarization $\zeta$ of a 2D Fermi liquid and the dashed curve is the boundary of the fully polarized liquid, both obtained from QMC simulations (see also fig. S17). We use the conduction band $g$ factor of 4.3, the dielectric constant, $\varepsilon = 4.6$, and the effective mass of electrons, $m_e^* = 0.7 m_0$ as parameters for the QMC simulations, where $m_0$ is the bare electron mass. Empty dots are onset densities of $\sigma-$ AP at each magnetic field. (**B**) Reduced spin susceptibility as a function of electron density obtained from QMC (dashed curve) and computed $\tilde{M}$ from the reflectance contrast spectra (blue dots). We conduct QMC simulations with the same values of the $g$ factor and effective mass as in panel (A) and use the dielectric constant as a free parameter. A good fit is obtained by using the dielectric constant, $\varepsilon = 4$, which deviates about 10% from the value used in panel (A). Inset: $\tilde{M}$ as a function of magnetic field for an electron density of 1.3, 1.8, and $2.0 \times 10^{12}$ cm$^{-2}$. (**C**) Scaled

magnetization curves ($\widetilde{M}/n$ as a function of $\mu_B H/k_B T$) in the Wigner crystal regime. The shapes and colors of markers represent different electron densities and temperatures. The black curve is a fit by Brillouin function with $g = 4.3$. (**D**) Reduced spin susceptibility ($\chi/\chi_0$) at $n = 0.28 \times 10^{12}$ cm$^{-2}$ as a function of electron temperature, showing Curie susceptibility.

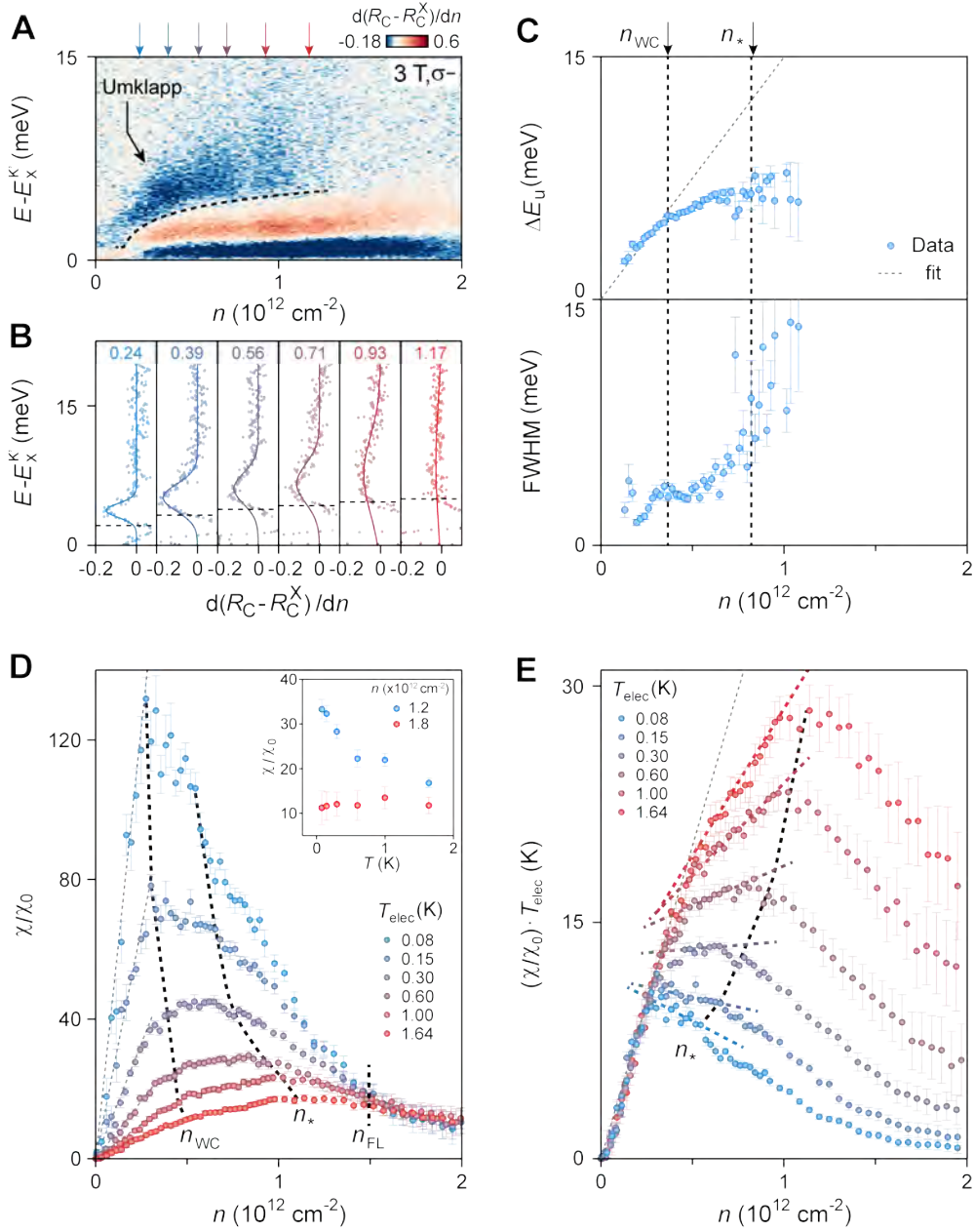

**Fig. 3. Quantum melting of a Wigner crystal.** (**A**) Color map of the derivative of reflectance contrast spectra (3 T, σ−), plotted against electron density and the energy detuning from the main exciton/RP resonance. A Lorentzian-fitted exciton/RP peak was subtracted to emphasize the higher-energy feature from umklapp scattering. (**B**) Cross-sections through the 2D map indicated by colored arrows in (A) at a fixed electron density, denoted by numbers in units of $10^{12}$ cm$^{-2}$. Each spectrum is fit with a Gaussian function, represented by colored solid lines. Black dashed lines in panel (A and B) mark the energy above which we used for fitting to avoid contributions from the residual exciton peak after subtraction. (**C**) The top panel presents energy difference

between exciton and umklapp features as a function of electron density. Gray dashed line is a linear fit assuming all the electrons are crystalized into a triangular lattice which provides the umklapp scattering momentum for excitons. From the slope $\Delta E_u/n = h^2/\sqrt{3}m_X$, we extract the exciton mass, $m_X = (1.15 \pm 0.05)m_0$. The bottom panel shows the full width at half maximum of umklapp scattering as a function of electron density. The black arrows on the top x-axis and black dashed lines indicate the characteristic densities, $n_{WC}$ and $n_*$. (**D**) Reduced spin susceptibility as a function of electron density at different temperatures. Gray dashed lines are Curie susceptibilities in the Wigner crystal regime. Black dashed lines indicate the density $n_{WC}(T)$ at which the spin susceptibility deviates from Curie (left), the density $n_*(T)$ at which the slope changes (right), and the density $n_{FL}$ above which the spin susceptibility does not exhibit temperature dependence. Inset, reduced spin susceptibility as a function of electron temperature at selected densities in the higher density regime. (**E**) The product of reduced spin susceptibility and temperature as a function of electron density. The Curie susceptibility is depicted by a gray dashed line. At low densities, the data points at different temperatures align with the Curie susceptibility. The colored dashed lines correspond to the linear interpolation of the susceptibility data between the density near $n_{WC}(T)$ and $n_*(T)$, obeying the lever rule. Deviations from the linear interpolation can be found in the vicinity of $n_{WC}(T)$ and $n_*(T)$ (see also fig. S14).

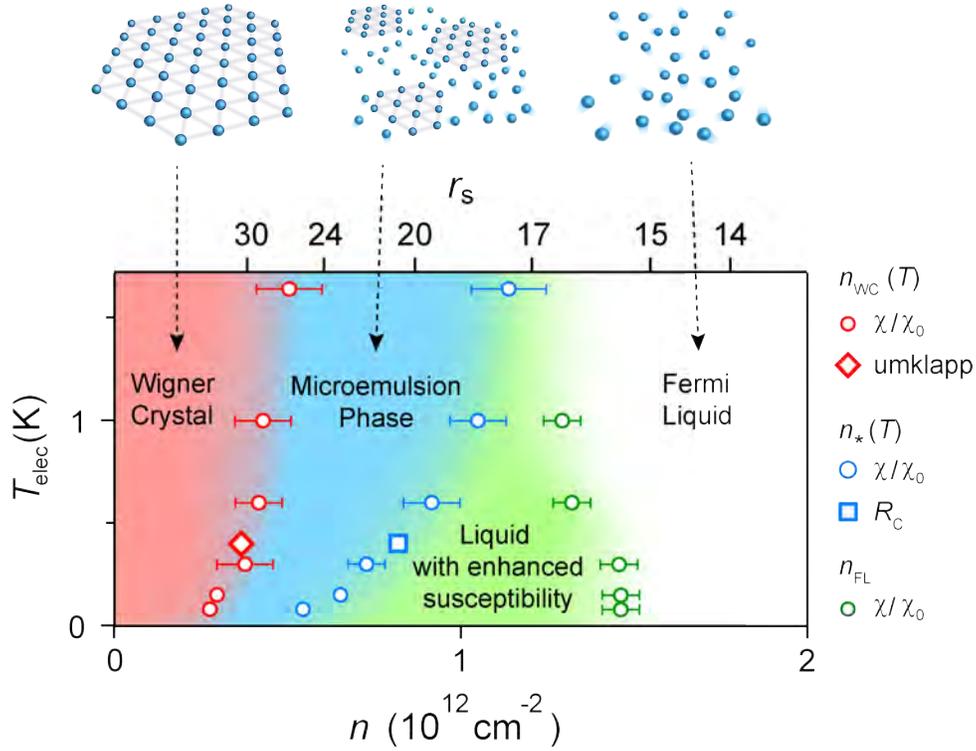

**Fig. 4. Phase diagram as a function of electron density and temperature.** Phase boundaries are obtained by the following characteristic densities: $n_{\text{WC}}(T)$, $n_*(T)$ and $n_{FL}$ from the spin susceptibility measurements (circles), $n_{\text{WC}}(T)$ from anomalies in umklapp scattering (diamond), and $n_*(T)$ from anomalies in reflectance spectra (square). The error bars associated with $n_{\text{WC}}(T)$ and $n_*(T)$ from the spin susceptibility measurements represent the density ranges that show deviations from the lever rule. The density $n_{FL}$ is defined as that at which the spin susceptibility deviates from the highest temperature (1.64 K) spin susceptibility, and the error bars indicate the nearest data points from the deviation. See also fig. S18 for the phase diagram that includes the density-dependent Fermi temperature of the non-interacting 2DEG.

**Materials and Methods**

Device fabrication and operation

Monolayer MoSe$_2$ and hBN flakes were exfoliated from bulk crystals onto 285 nm SiO$_2$/Si substrates. MoSe$_2$ monolayers were identified by an optical microscope. Thicknesses of hBN flakes were measured by atomic force microscopy. The flakes were stacked by the dry transfer method using a polydimethylsiloxane stamp and a thin layer of polycarbonate. The stacked heterostructure was then transferred onto the pre-patterned bottom gate, which consists of a 1 nm Cr layer and a 9 nm Pd/Au alloy layer fabricated using electron-beam lithography and thermal evaporation. Electrical contacts to the MoSe$_2$ layer and the bottom gate consisting of 5 nm Cr layer and 90 nm Au layer were deposited via thermal evaporation.

To dope the monolayer MoSe$_2$ at dilution refrigerator temperatures, we grounded the contacts to MoSe$_2$, applied a voltage $V_{bg}$ to the back gate and at the same time illuminated the whole sample with a broadband light to activate the contacts. After doping was finished, the activation light was removed, and the sample was thermalized for 0.5 s. The optical measurements were performed after the thermalization was finished. The doping onset voltage $V_0$ is determined as the voltage at which the reflectance contrast spectra or magneto-optical signal starts to deviate from the neutral regime. The electron density is calculated from the parallel-plate capacitor model $n = \varepsilon_0 \varepsilon_r (V_{bg} - V_0)/d_{bg}$, where $\varepsilon_0$ is the vacuum permittivity, $\varepsilon_r \approx 3.9$ is the dielectric constant of the hBN, and $d_{bg}$ is the thickness of the bottom hBN dielectric.

Optical measurements

Reflectance spectroscopy and magneto-optical measurements were performed with a home-built scanning confocal microscope based on a dilution refrigerator (Bluefors) with which the sample lattice temperature can reach 16 mK. The sample is mounted in the center of a superconducting magnet (AMI) capable of applying ±9 T perpendicular magnetic field. A piezo-electric stage (attocube) was used to precisely position the sample. The microscope consists of an apochromatic cryogenic objective (attocube LT-APO/LWD, NA 0.65), two fused silica plano-convex lenses (OptoSigma) at around 4 K and 50 K, two achromatic doublet lenses (Thorlabs) and a galvo scanner (Thorlabs). A compensated full-wave liquid crystal retarder (Thorlabs LCC1413-B) was placed in the shared light path of incoming and outcoming beams to impose the same $\pm\lambda/4$ retardance to both beams without mechanical movement. A polarimeter (Thorlabs PAX1000IR1) was used to confirm the circular polarization outside of the dilution refrigerator. To ensure a polarization-independent light path inside the dilution refrigerator, we examined the reflection isotropy of a bare $SiO_2/Si$ substrate mounted inside with a 360° rotation of the linearly polarized light.

For reflectance spectroscopy, a tungsten-halogen lamp (Thorlabs SLS201L) filtered to 720 ~ 800 nm range was coupled to a single mode fiber, collimated with an objective (Olympus PLN 10×, NA 0.25) and directed to the sample, creating a diffraction-limited spot. The light power on the sample was variable but always kept below 0.7 nW. The reflected light was collected by a spectrometer with a 1200 g/mm grating and a CCD camera (Princeton Instruments BLAZE). For magneto-optical measurements, we switched to a continuous-wave ultra-narrow linewidth Ti:Sapphire laser (M Squared SOLSTIS) and limited light power on the sample to 60 fW (unless otherwise noted). The reflected laser was collected directly by a CCD camera (Princeton Instruments BLAZE).

Magneto-optical measurements

Magneto-optical measurements using a single energy excitation are conducted at a fixed energy (1.636 eV) within a small magnetic field range of -1 T to 1 T for electron magnetization and spin susceptibility measurements in Fig. 2, C and D and Fig. 3, D and E. Since the bare exciton Zeeman splitting is small within this range of magnetic fields ($\leq 0.1$ meV for each $\sigma\pm$ polarization), its contribution to the signal is small, which can be subsequently subtracted. We also note that $I_{max}^{\sigma+} \approx I_{max}^{\sigma-}$ (reflection spectrum at a density of $1.0\times10^{13}$ cm$^{-2}$) within -1 T to 1 T, and $\widetilde{M} = (I^{\sigma+}/I_{max}^{\sigma+} - I^{\sigma-}/I_{max}^{\sigma-})(I^{\sigma+}/I_{max}^{\sigma+} + I^{\sigma-}/I_{max}^{\sigma-}) \approx (I^{\sigma+} - I^{\sigma-})/(I^{\sigma+} + I^{\sigma-})$. We can thus use the reflected intensity, instead of the normalized one to obtain $\widetilde{M}$. To convert the measured magneto-optical signal $\widetilde{M} = \frac{I^{\sigma+} - I^{\sigma-}}{I^{\sigma+} + I^{\sigma-}}$ to magnetization $M$, we introduce a density dependent scaling factor that is field independent: $\widetilde{M} = A(n)M$. At high fields when the system is fully polarized, the magnetization is known as $M_S = \frac{1}{2}g\mu_B n$, so that the scaling factor can be extracted $A(n) =$ $\frac{2\widetilde{M}_S}{g\mu_B n}$. Therefore, at an arbitrary field, the magnetization and susceptibility are $M = \frac{\widetilde{M}}{\widetilde{M}_S} \cdot \frac{1}{2} g\mu_B n$ and $\chi = \frac{d\widetilde{M}}{dH} \cdot \frac{1}{\widetilde{M}_S} \cdot \frac{1}{2} g\mu_B n$. The spin susceptibility of the non-interacting 2DEG is $\chi_0 = \left(\frac{g\mu_B}{2}\right)^2 \frac{m^*}{\pi\hbar^2}$. Thus, the reduced spin susceptibility can be expressed as $\chi/\chi_0 = \frac{d\widetilde{M}}{dH} \cdot \frac{1}{\widetilde{M}_S} \cdot \frac{2\pi\hbar^2 n}{g\mu_B m^*}$. We use the conduction band $g$ factor value of 4.3 (determined by fitting the measured $\widetilde{M}(n,H)$ using the Brillouin function at $n < n_{WC}$) and the effective mass of electrons, $m_e^* = 0.7 m_0$.

**Supplementary Text**

Determination of repulsive polaron (RP) and attractive polaron (AP) resonances

Due to the layered structure of the sample, the optical susceptibility near the excitonic resonance in a MoSe$_2$ monolayer is represented as phase shifted quantity,

$$\Psi(E) = e^{i\alpha}\chi(E),$$

where $\chi(E)$ denotes the optical susceptibility of a MoSe$_2$ monolayer, $\alpha$ denotes the phase factor (we approximate $\alpha$ by a constant $\alpha_0$ in the energy range of interest), and $E$ is the photon energy. In practice, measured reflectance contrast spectra, $R_C(E)$ correspond to the imaginary part of $\Psi(E)$. Thus, we can express reflectance contrast as,

$$R_C(E) = \Psi''(E) = \chi''(E)\cos\alpha_0 + \chi'(E)\sin\alpha_0,$$

where $\chi'(E)$ denotes the real part and $\chi''(E)$ represents the imaginary part of $\chi(E)$.

Since the optical susceptibility $\chi(E)$ is well approximated by the resonance form:

$$\chi(E) = -\frac{A^2}{(E-E_0)+i\gamma/2},$$

each RP and AP resonance in the reflectance contrast can be described by the following equation.

$$R_{RP,AP}(E) = \frac{A^2}{(E-E_0)^2+\gamma^2/4}\left[\frac{\gamma}{2}\cos\alpha - (E-E_0)\sin\alpha\right] + C,$$

where $A^2$, $E_0$, and $\gamma$ denotes the amplitude, energy, and linewidth of the RP or AP resonance and $C$ is constant background.

Reflectance contrast is defined as $R_C^*(E) = \frac{I(E)}{I_\infty(E)} - 1$, where $I(E)$ is the reflected light intensity and $I_\infty(E)$ is the reflected light intensity at an electron density that is high enough to bleach all excitonic resonances. In practice, electron doping of a MoSe$_2$ monolayer can be achieved up to a density of $1.0 \times 10^{13}$ cm$^{-2}$ by applying a bottom gate voltage while keeping the gate-sample current below 1 nA. At this electron density, the reflection spectrum, $I_{max}(E)$ exhibits negligible RP resonances, while the AP resonances are still observable. Therefore, Lorentzians give a good fit for RP resonances in reflectance contrast spectra, $R_C(E) = \frac{I(E)}{I_{max}(E)} - 1$, but the same fitting approach is not readily applicable for analysing the AP resonances. To address this, we determine the finite AP resonance in the reflectance contrast spectrum at an electron density of $1.0 \times 10^{13}$ cm$^{-2}$ by Lorentzian fitting the obtained $R_{Cmax}^*(E) = \frac{I_{max}(E)}{I_\infty(E)} - 1$, and subtracting this fitted equation from all other reflectance contrast spectra for lower densities.

We note that the reflectance contrast spectra can be expressed as,

$$R_C(E) = \frac{I(E)}{I_{max}(E)} - 1 = \frac{I(E)/I_\infty(E)}{I_{max}(E)/I_\infty(E)} - 1 = \frac{1+R_C^*(E)}{1+R_{Cmax}^*(E)} - 1,$$

where $R_C^*(E) = \frac{I(E)}{I_\infty(E)} - 1$, represents the reflectance contrast spectra obtained in the ideal case. Since the reflectance contrast from the AP resonance at an electron density of $1.0 \times 10^{13}$ cm$^{-2}$ is small ($R_{Cmax}^*(E) \ll 1$), $R_C(E)$ can be approximated as,

$$R_C(E) \approx (1 + R_C^*(E))(1 - R_{Cmax}^*(E)) - 1 = R_C^*(E) - (1 + R_C^*(E))R_{Cmax}^*(E)$$

Consequently, we can determine $R_{Cmax}^*(E) = \frac{1}{1+R_C^*(E)}(R_C^*(E) - R_C(E))$.

When a MoSe$_2$ monolayer is intrinsic, the fitted Lorentzian function for the main excitonic resonances, $R_C^X(n=0, E)$ is close to $R_C^*(n=0, E)$ since there is no AP resonance (see Fig. S1A for the fitted Lorentzian, $R_C^X(n=0, E)$ with reflectance contrast spectrum $R_C(n=0, E)$ in the neutral regime). Next, we fit this background AP resonance in the obtained $R_{Cmax}^*(E)$ ($\approx R_C^X(n=0, E) - R_C(n=0, E)$) as shown in Fig. S1B, and subsequently subtract it out for reflectance contrast spectra at all other electron densities. In Fig. S1C and S1D, we show the AP resonances in the reflectance contrast spectra at various electron densities after background subtraction. Additionally, we plot color maps in Fig. S2, showing left (σ−) and right (σ+) circularly polarized reflectance contrast spectra at 9 T after and before the background subtraction. From the background subtracted spectra, we extract the AP amplitude, energy, and linewidth by Lorentzian fitting at different magnetic fields. The onset density of σ− AP resonance at each field is determined from a linear fitting to the σ− AP amplitude as a function of density (Fig. S3).

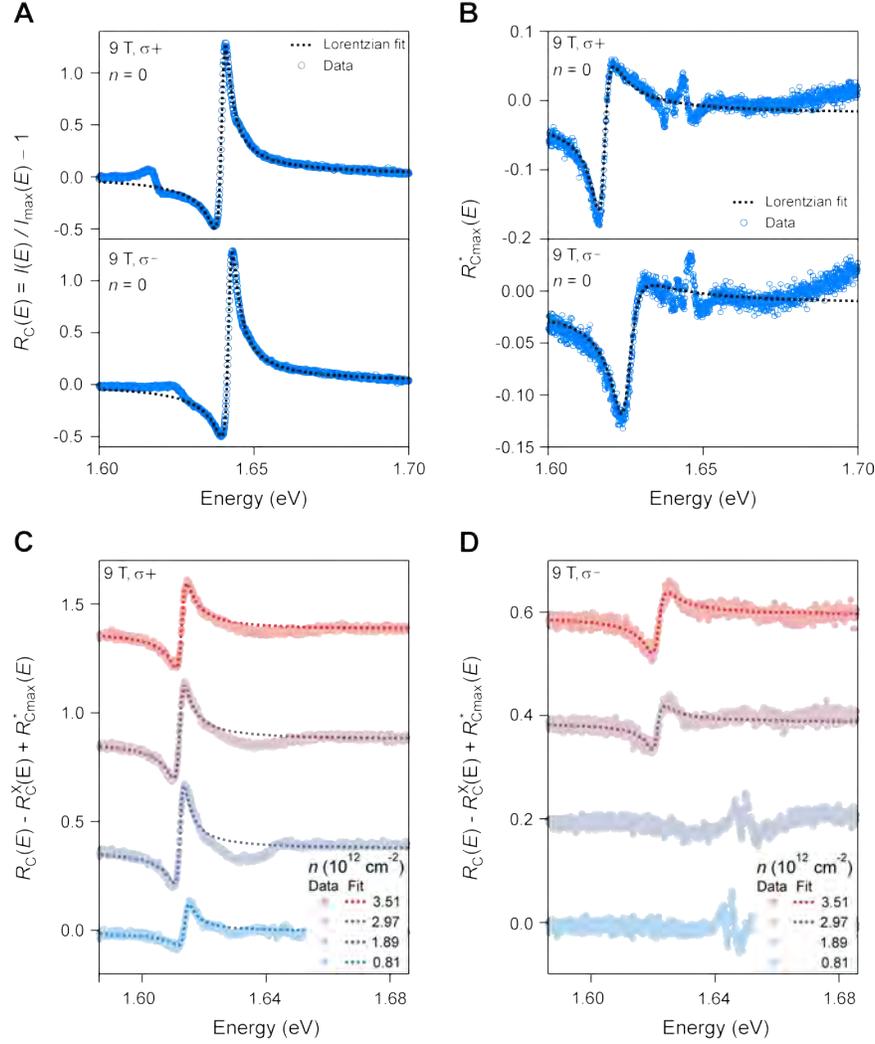

**Fig. S1. Lorentzian fitting of AP resonances at 9T.** (A) The reflectance contrast spectrum $R_C(E)$, in the neutral regime at 9T is shown for right (σ+, top panel) and left (σ−, bottom panel) circular polarization. The corresponding fitted Lorentzian function, $R_C^X(E)$, is also shown. (B) The obtained reflectance contrast spectrum for right (σ+, top panel) and left (σ−, bottom panel) circular polarization at an electron density of $1.0 \times 10^{13}$ cm$^{-2}$, $R_{Cmax}^*(E)$, is shown with the fitted Lorentzian function for the background AP resonance. (C) Right (σ+) circularly polarized reflectance contrast spectrum showing predominantly the AP resonances at various electron densities after the background AP resonance subtraction and subtraction of $R_C^X(E)$. The spectra and the fit are vertically shifted for clarity. (D) Left (σ+) circularly polarized reflectance contrast spectrum showing predominantly the AP resonances at various electron densities after the background subtraction and subtraction of $R_C^X(E)$. At electron densities of $0.81 \times 10^{12}$ cm$^{-2}$ and $1.89 \times 10^{12}$ cm$^{-2}$, the AP resonances are negligible, resulting in the Lorentzian fit being not applicable. The spectra and the fit are vertically shifted for clarity.

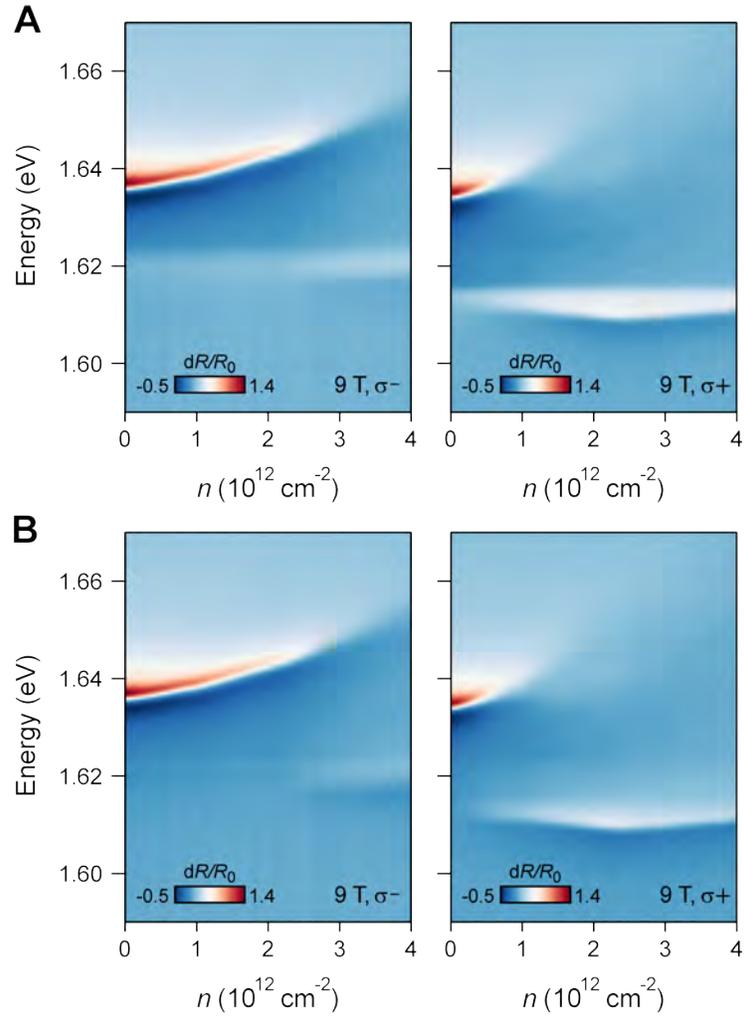

**Fig. S2. Reflectance contrast spectra at 9T as a function of electron density.** Color maps for the left (σ−) and right (σ+) circularly polarized reflectance contrast spectra (A) before and (B) after the background AP resonance subtraction.

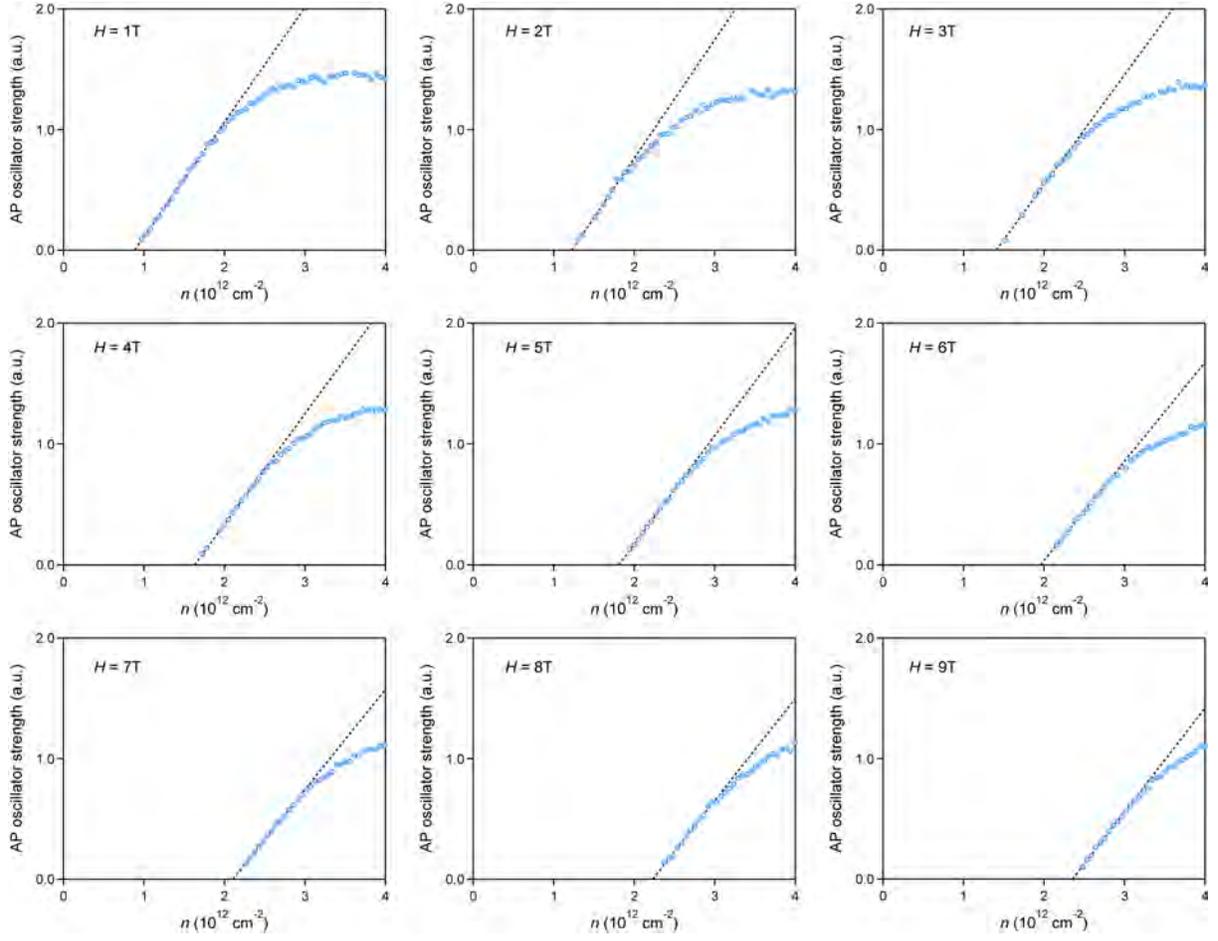

**Fig. S3. Determination of onset density of the σ- AP resonance.** We extract σ- AP amplitude by Lorentzian fitting and plot it as a function of electron density. Employing a linear fit to the AP amplitude allowed us to determine the onset electron density of the AP resonance at each magnetic field.

Critical density between fully and partially spin/valley polarized liquid

The critical density between fully and partially spin/valley polarized liquid for a non-interacting 2DEG can be expressed by equating the Fermi energy to the Zeeman energy. This critical density is given by,

$$n_c = (\mu_B/2\pi\hbar^2)\, gm^*H,$$

where $\mu_B$ represents the Bohr magneton, $g$ denotes the g factor, and $m^*$ is the electron effective mass. We use conduction band g factor of 4.3 that we extract from the fit using the Brillouin function to the magneto-optical signal, $\widetilde{M}$ (see the SI5) and effective mass $m^*$ value of 0.7 (*58*). The calculated $n_c$ and the onset densities of σ− AP at different magnetic fields are plotted in Fig. S4. We note that the observed critical densities from the σ− AP resonance exceed the value for a non-interacting 2DEG by an order of magnitude, providing evidence of the strong Coulomb interactions in a MoSe$_2$ monolayer.

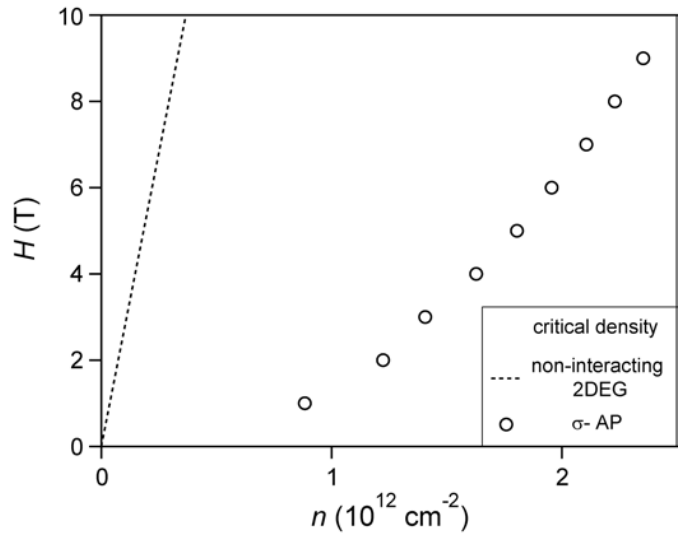

**Fig. S4. Critical density between fully and partially spin/valley polarized liquid.** We plot extracted onset electron density of the σ- AP resonance together with calculated critical density for a non-interacting 2DEG as a function magnetic field.

Spin susceptibility from magneto-optical signal $\widetilde{M}$

To validate the magneto-optical signal $\widetilde{M} = (I^{\sigma+}/I^{\sigma+}_{max} - I^{\sigma-}/I^{\sigma-}_{max})/(I^{\sigma+}/I^{\sigma+}_{max} + I^{\sigma-}/I^{\sigma-}_{max})$ as a probe for the electron magnetization, we obtain signal $\widetilde{M}$ from the reflectance contrast spectra and plot a ($n, H$) phase diagram of the 2D electron with respect to electron density and magnetic field (Fig. S5). We find that constructed phase diagram exhibits transition from fully polarized to partially polarized Fermi liquid, which agrees well with the σ− AP onset densities.

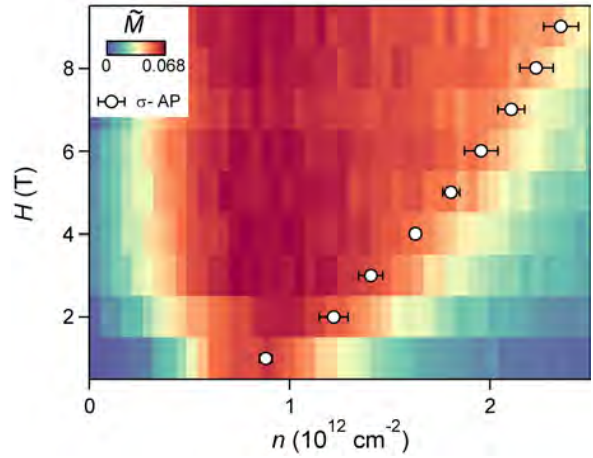

**Fig. S5. Magneto-optical signal $\widetilde{M}$ from the reflectance contrast spectra.** Phase diagram of the 2D electron system with respect to electron density and magnetic field constructed by using the magneto-optical signal $\widetilde{M}$. Empty dots are onset electron densities of σ− AP at each magnetic field.

To account for the bare exciton Zeeman splitting in the intrinsic regime at each magnetic field, we take normalized reflection ($I^{\sigma\pm}/I^{\sigma\pm}_{max}$) at fixed detuning energies denoted as $\Delta E = E(H) - E_X^{\sigma\pm}(H)$, where $E(H)$ is the selected energy for the analysis, $E_X^{\sigma\pm}(H)$ is the magnetic field dependent exciton resonance energy at zero doping, and $\Delta E$ is the fixed detuning energy. We take $\Delta E = $ -5 meV, red detuned from the exciton resonance at zero doping to ensure that the main contribution is from the real (dispersive) parts of the optical susceptibility. To see this, we analyse the reflectance contrast spectra to separate the real and imaginary parts of the optical susceptibility and compare the contribution from each at selected energy, $E(H)$. In the previous section, we note that the reflectance contrast spectra can be expressed as, $R_C(E) = \Psi''(E) = \chi''(E)\cos\alpha_0 + \chi'(E)\sin\alpha_0$, where $\chi'(E)$ denotes the real part and $\chi''(E)$ represents the imaginary part of $\chi(E)$.

With the approximation for the optical susceptibility near the excitonic resonance by the Lorentzian function, we can express the reflectance contrast as,

$$R_{RP,AP}(E) = \frac{A^2}{(E-E_0)^2 + \gamma^2/4}\left[\frac{\gamma}{2}\cos\alpha - (E-E_0)\sin\alpha\right] + C$$

Thus, by fitting the reflectance contrast spectra using this Lorentzian equation, we can extract $\chi'(E)\sin\alpha_0$ and $\chi''(E)\cos\alpha_0$. The fitted results for the left (σ−) and right (σ+) circularly polarized reflectance contrast spectra measured at 1T are shown in Fig. S6. We see that the signal from $\chi'(E)\sin\alpha_0$ dominates over that from $\chi''(E)\cos\alpha_0$ at $E(1T) = 1.636$ eV ($\Delta E$ = -5 meV). If we take positive detuning energies to compute $\widetilde{M}$, the real and imaginary part of $\chi(E)$ are mixed, resulting in sign change for $\widetilde{M}$ as a function of electron density. This is due to a spectral blueshift of the RP resonance.

A simple calculation shows that the magneto-optical signal is proportional to polar Kerr rotation angle or ellipticity.

$$\widetilde{M} = \frac{(I^{\sigma+}/I_{max}^{\sigma+} - I^{\sigma-}/I_{max}^{\sigma-})}{(I^{\sigma+}/I_{max}^{\sigma+} + I^{\sigma-}/I_{max}^{\sigma-})} \approx \frac{(\chi'_{\sigma+}\sin\alpha_{\sigma+} - \chi'_{\sigma-}\sin\alpha_{\sigma-})}{(\chi'_{\sigma+}\sin\alpha_{\sigma+} + \chi'_{\sigma-}\sin\alpha_{\sigma-} + 2)} \quad (1)$$

Here we use $\frac{I(E)}{I_{max}(E)} - 1 \approx \Psi''(E) = \chi''(E)\cos\alpha_0 + \chi'(E)\sin\alpha_0$, and the observation that the signal from $\chi'(E)\sin\alpha_0$ dominates over that from $\chi''(E)\cos\alpha_0$ at 1.636 eV.

At low magnetic fields, $\sin\alpha_{\sigma+} \approx \sin\alpha_{\sigma-}$ for various electron densities, and we have

$$\widetilde{M} \approx \frac{(\chi'_{\sigma+} - \chi'_{\sigma-})}{(\chi'_{\sigma+} + \chi'_{\sigma-} + C)} \approx -\frac{4\pi}{((n^2)' - 1 + 2\pi C)}\sigma_{xy}'/\omega \quad (2)$$

Where $C = 2/\sin\alpha_0$, $n$ is refraction index in the absence of magneto-optical effect and $\sigma_{xy}$ is the off-diagonal part of the ac conductivity, which is the leading term in any magneto-optic effects. The equality in (2) can be found in refs. (*59, 60*).

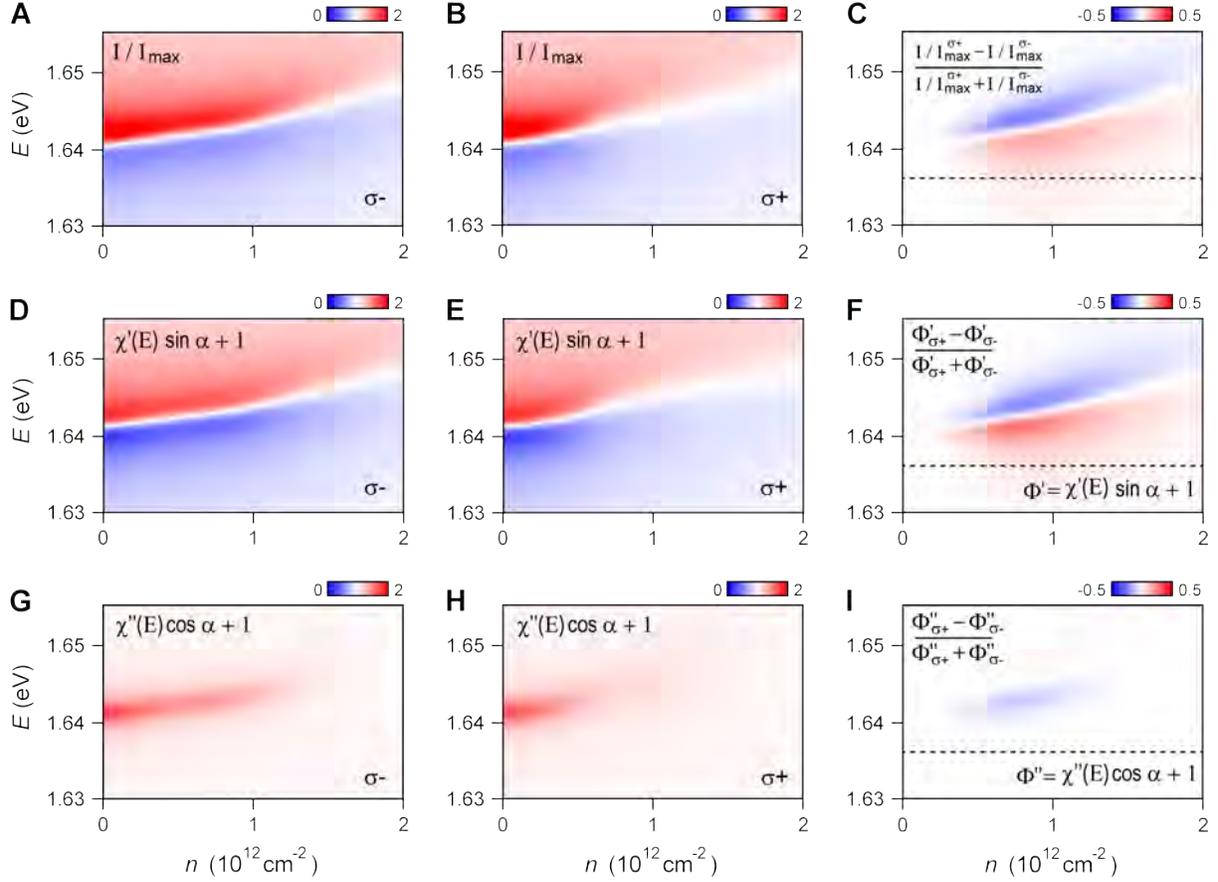

**Fig. S6. Decomposition of fitted Lorentzian spectral profile into real and imaginary parts of the optical susceptibility and their contribution to signal $\widetilde{M}$.** The fitted results for the (**A**) left and (**B**) right circularly polarized reflectance contrast spectra measured at 1T are presented. Additionally, the signal $\widetilde{M}$, which represents the difference in normalized reflected intensity between σ± light divided by their sum is shown in (**C**). The decomposed (**D**, **E**) real and (**G**, **H**) imaginary parts of the optical susceptibility are shown, along with the difference over the sum in (**F**) and (**I**), respectively. Selected detuning energy, $\Delta E$ = -5 meV (1.636 eV at 1T) is indicated by black dashed lines in (C, F, I). It is shown that the signal from $\chi'(E)\sin\alpha_0$ dominates over that from $\chi''(E)\cos\alpha_0$ at $\Delta E$ = -5 meV (1.636 eV).

Spin susceptibility from various observables

We extract the spin susceptibility from various observables and compare them to that obtained by the magneto-optical signal $\widetilde{M}$.

(1) We use AP amplitude difference between σ± polarized reflectance spectra ($\Delta \text{Amp}_{AP,\pm}$) as a probe for the electron magnetization. Since the AP resonance arises from the inter-valley exciton-electron interaction, the AP amplitude in σ+ (σ−) polarized reflectance spectra is proportional to spin down (up) electrons in the K' valley (K valley). To convert the AP amplitude difference to magnetization $M$, we use the same method as we use for the magneto-optical signal $\widetilde{M}$. We introduce a density dependent scaling factor $\Delta \text{Amp}_{AP,\pm} = A(n) \cdot M$. At high fields (9 T in our study) when the system is fully polarized, the magnetization is known as $M_S = \frac{1}{2} g \mu_B n$, so that the scaling factor can be extracted $A(n) = \frac{2 \cdot (\Delta \text{Amp}_{AP,\pm} \text{ at 9 T})}{g \mu_B n}$. Fig. S7A shows the AP amplitude difference as a function of magnetic field at various electron densities, and we can obtain the spin susceptibility by a linear fit at each density.

(2) We use exciton/RP Zeeman energy splitting ($\Delta E_{RP,\pm} = E_{RP,\sigma+} - E_{RP,\sigma-}$) as a probe for the electron magnetization. In a phenomenological model based on a mean-field interaction, excitons sense the magnetic moment of the electron spins, giving rise to Zeeman energy splitting between right and left circularly polarized reflectance spectra (*32*). We exclude to the contribution from the bare exciton Zeeman energy splitting by subtraction. Again, we use the same method to convert ($\Delta E_{RP,\pm}$ to magnetization $M$. Fig. S7B shows the exciton/RP Zeeman energy splitting as a function of magnetic field at various electron densities, from which we extract the spin susceptibility at each density.

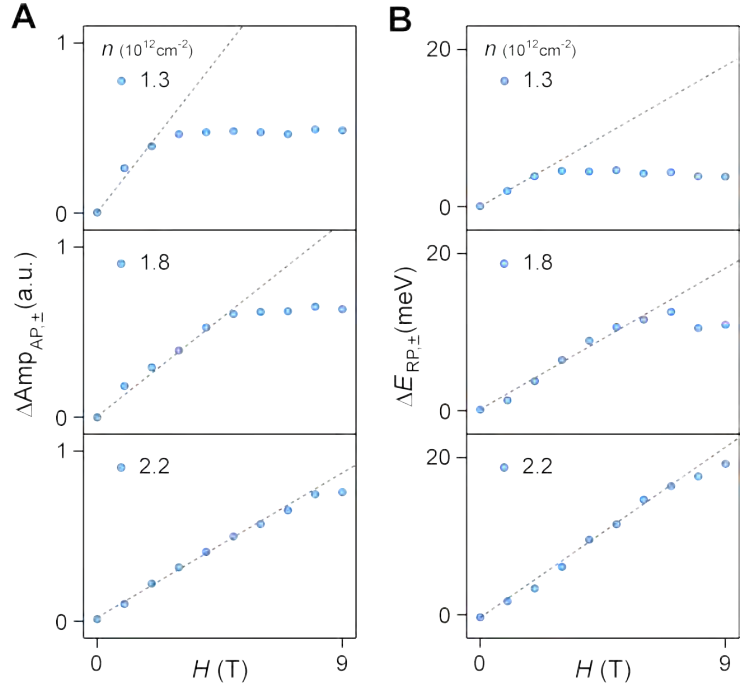

**Fig. S7. (A) AP amplitude difference and (B) exciton/RP Zeeman energy splitting as a function of magnetic field at various electron densities.**

Fig. S8 shows the spin susceptibility extracted from method (1) & (2), $\widetilde{M}$ computed from the reflection spectra (Fig. 2B in main text) and $\widetilde{M}$ measured using a single energy excitation (Fig. 3D in main text). The results from the different methods are in good agreement with QMC simulations based on material parameters using the conduction band *g* factor of 4.3, the dielectric constant, $\varepsilon = 4$, and the effective mass of electrons, $m_e^* = 0.7 m_0$ (the same as we use in Fig. 2B in main text).

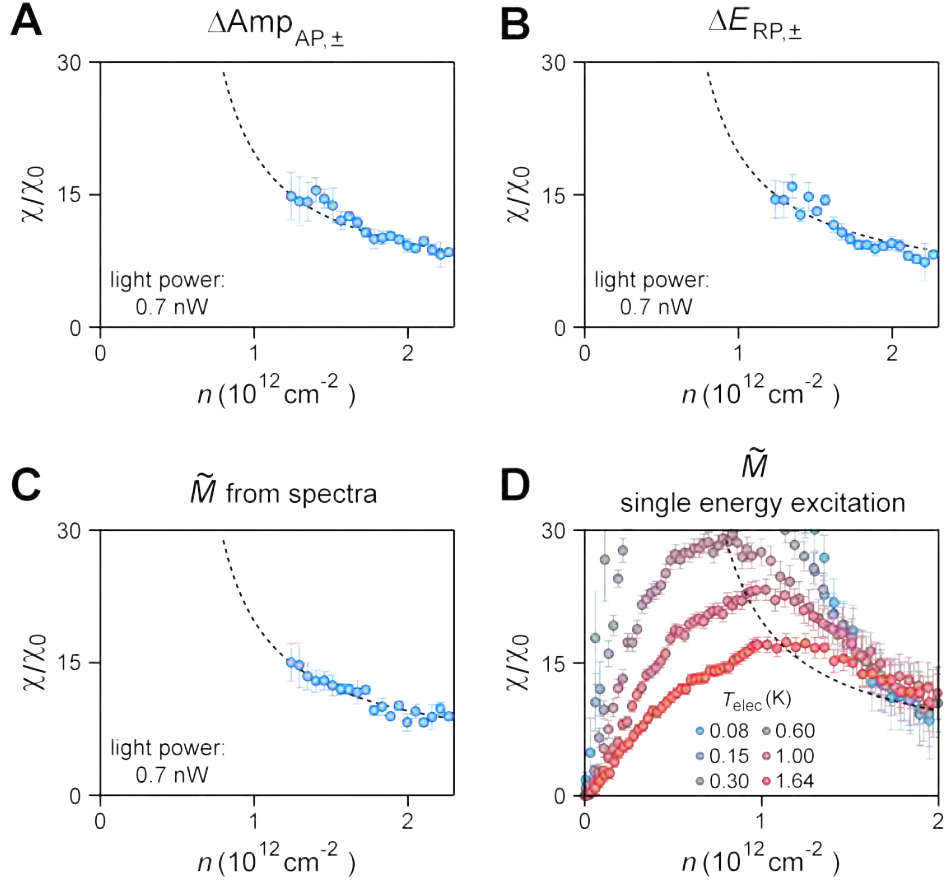

**Fig. S8.** Reduced spin susceptibility extracted from (**A**) AP amplitude difference (**B**) exciton/RP Zeeman energy splitting (**C**) $\widetilde{M}$ computed from the spectra in a narrow spectral range centered at 1.636 eV and (**D**) $\widetilde{M}$ measured using a single energy excitation (1.636 eV) at 60 fW incident power on the sample.

Estimation of the electron temperature and the conduction band g-factor

We plot the normalized $\widetilde{M}(H)$ curves at the electron density of $0.30 \times 10^{12}$ cm$^{-2}$ under varying light powers at the base lattice temperature of 16 mK (Fig. S9). From a fit using the Brillouin function, we estimate the electron temperature with conduction band g-factor of 4.3. The estimation of the conduction band g-factor is discussed futher below.

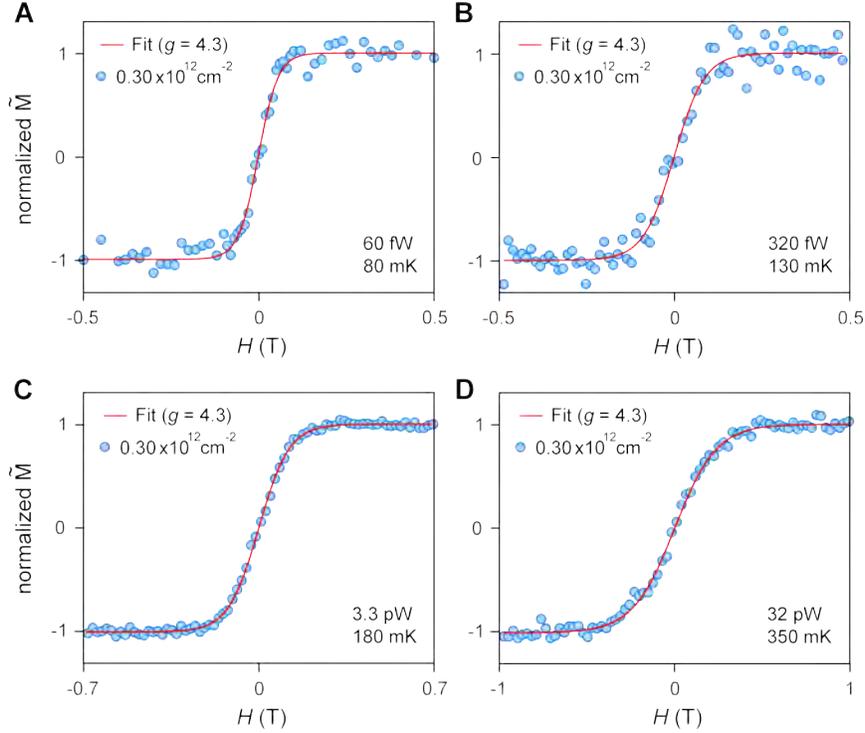

**Fig. S9. Electron temperature estimation at low density under different incident light powers at the base lattice temperature of 16 mK.** (**A**) 60 fW, (**B**) 320 fW, (**C**) 3.3 pW, and (**D**) 32 pW. The estimated electron temperatures are: (**A**) 80mK, (**B**) 130 mK, (**C**) 180 mK, and (**D**) 350 mK.

To see how electron temperatures change at given lattice temperatures under fixed 60 fW light power, we plot the $\widetilde{M}(H)/n$ curves measured under fixed 60 fW light power as a function magnetic field, $H$, within the low-density regime, n ≤ $0.35 \times 10^{12}$ cm$^{-2}$. Then, we assumed that for the lattice temperature, $T_{lattice} \geq 300$ mK, the electron temperature, $T_{elec}$ is equal to the lattice temperature. Then, from a global fit using the Brillouin function to the $\widetilde{M}(H)/n$ curves at six different lattice temperatures (16 mK, 150 mK, 300 mK, 600mK, 1K, 1.64K), we obtain the g-factor of 4.3 for the conduction band in the absence of the interaction effects as well as the electron temperatures at low lattice temperatures, $T_{lattice} < 300$ mK. We note that at $T_{lattice}$ of 150 mK, we obtain $T_{elec}$ of 150 mK, indicating that the system cooling power is large enough to minimize the laser induced heating effect. At $T_{lattice}$ of 16 mK (base temperature), we obtain $T_{elec}$ of 80 mK (Fig. S10).

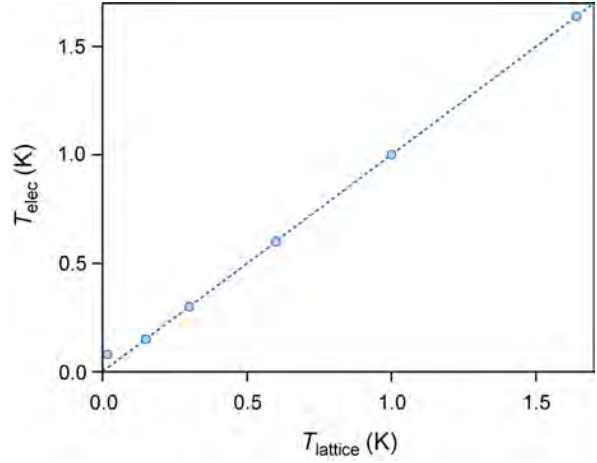

**Fig. S10. Estimated electron temperatures at the given lattice temperatures using the Brillouin function fit.**

One may worry the difference between the extracted $T_{elec}$ of 80 mK and $T_{lattice}$ of 16 mK could be due to an erroneous application of the Brillouin fit in a regime where the system has a propensity toward some form of magnetic order at low temperature, so that $T_{elec} \neq T_{lattice}$ would rather be an "effective temperature" taking into account effects of strong magnetic interactions (In this scenario, $T_{elec} > T_{lattice}$ would be more likely indicative of antiferromagnetism, but we consider in the following the more general case of $T_{elec}$ different from $T_{lattice}$). We can rule out such a scenario by considering the expected evolution of $M(H)$ in the two cases of low temperature ordered ferromagnetic and antiferromagnetic phases and argue the independent spin picture gives a better description of the data:

1) If there is a tendency toward ferromagnetism at low temperatures, this would manifest as a deviation in $M(H)$ from the simple hyperbolic tangent function. For instance, in mean-field theory, we can write, $M \propto \tanh[(H + M(H))/T]$, and we would find deviations in $M(H)$ from the simple $\tanh(H/T)$. Even above but near the ferromagnetic transition temperature, the hyperbolic tangent function cannot give a good fit to the magnetization curves. Thus, the high-quality $\tanh(H/T)$ fit shown in Fig. 2C strongly suggests that the system does not have a tendency toward ferromagnetism at the currently accessible temperatures. Of course, we cannot rule out ferromagnetic order at much lower temperatures than those we've accessed.

2) If there is a low-temperature tendency toward antiferromagnetism, the curves may remain well-fitted by $\tanh(H/T)$ (at least in mean-field theory, this is indeed the case above the Néel temperature). However, because the strength of the magnetic interactions is density dependent, the Néel temperature would also be density dependent. Thus, if the electron temperature we have accessed is indeed near the Néel temperature, we would expect our extracted electron temperature to depend on density. However, as shown in Fig. 2C, all the magnetization curves overlap for different densities (with the same extracted temperature), suggesting that the electron temperature is essentially density independent. Thus, tendency toward antiferromagnetic order is also unlikely at the currently accessible temperatures. As in the ferromagnetic case, we cannot rule out antiferromagnetic order at much lower temperatures.

Determination of umklapp peak energy and widths

The umklapp scattering properties of the exciton are obtained from reflection spectra. The isolation of the umklapp features proceeds in two steps: We first subtract the fitted RP lineshape for each electron density (see Fig. S1 and S2 for the fitting procedure). Then, as a second step, we differentiate the subtracted spectrum with respect to electron density to further highlight the umklapp feature. The associated lineshape of the subtracted and differentiated spectrum is more complex than that of the undifferentiated one, and we have thus not attempted to fit it to any particular model spectrum. Instead, we identify the peak position of the high energy umklapp resonance via a simple Gaussian fit. Since the residual exciton peak after subtraction is of similar intensity to the umklapp feature, we defined a density dependent fitting range as shown in Fig. S11. Each spectrum at a given density is shown with a Gaussian fit in Fig. S12. The extracted fitting parameters are plotted in Fig. 3C, where the error bars are fitting errors.

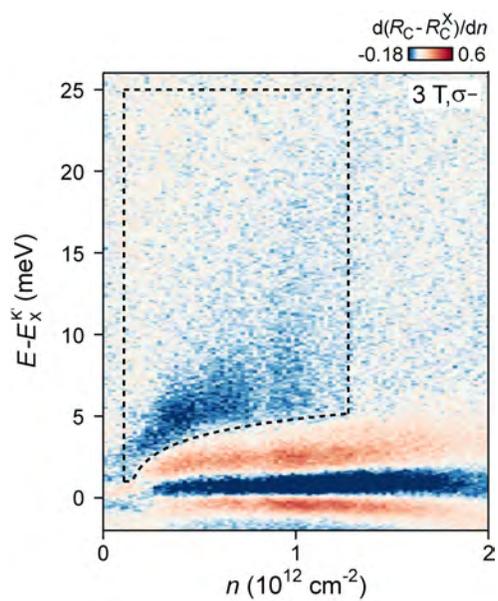

**Fig. S11. Range for the Gaussian fitting.** Derivatives of reflectance contrast with respect to electron density, with Lorentzian-fitted exciton peak subtracted. Black dashed lines mark the range for umklapp peak Gaussian fitting.

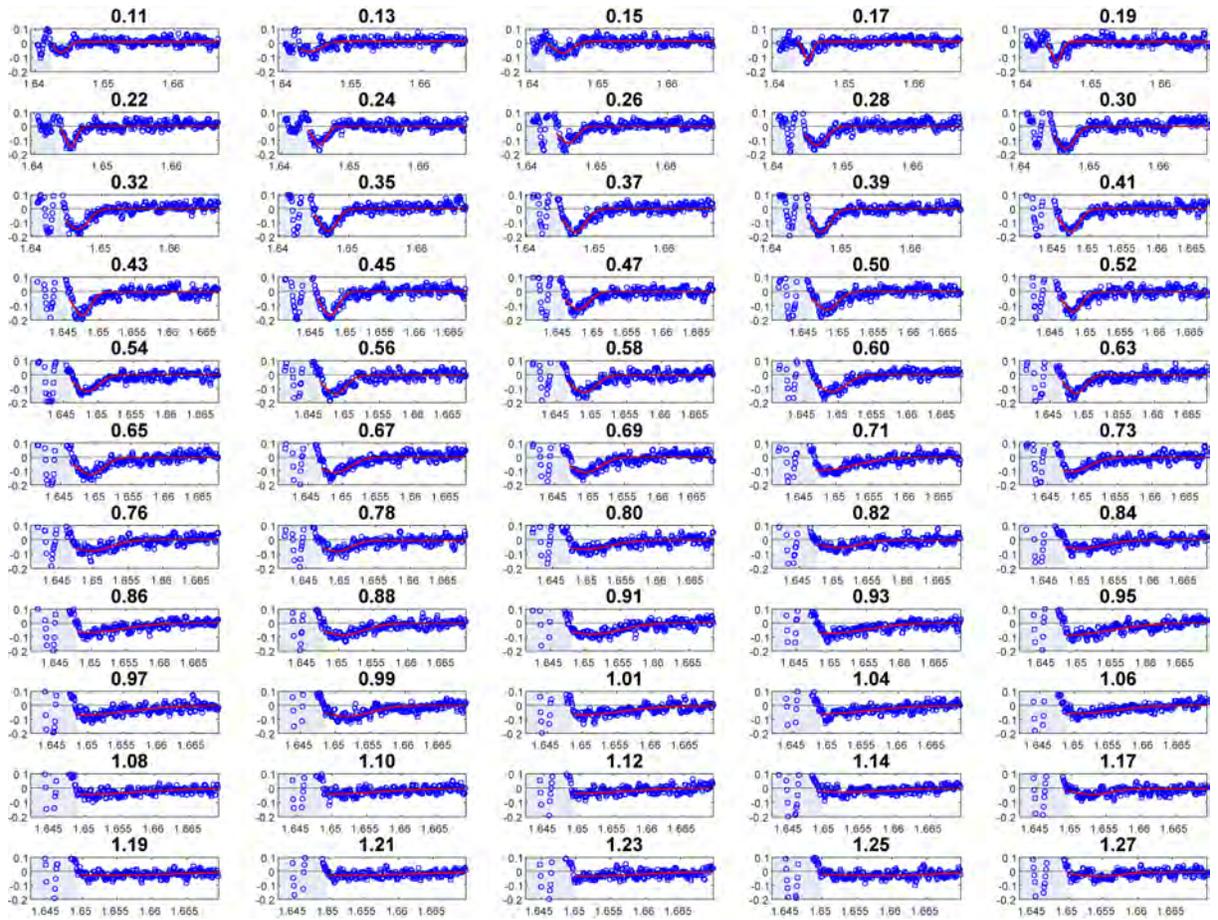

**Fig. S12. Gaussian fitting to the subtracted and differentiated spectrum.** Blue dots are experimental data. Red curves are gaussian fits. Light gray shades mark the energy range excluded for fitting. The title of each figure is electron density in the unit of $10^{12}$ cm$^{-2}$.

Anomalies in the main excitonic properties

Near the density of $n_* = 0.9 \times 10^{12}$ cm$^{-2}$, a noticeable discontinuity of the exciton/RP resonance energy with respect to electron density, $dE_X/dn$ is observed, signifying a change in the slope of the exciton/RP resonance energy. A pronounced decrease in the oscillator strength is also observed near $n_*$.

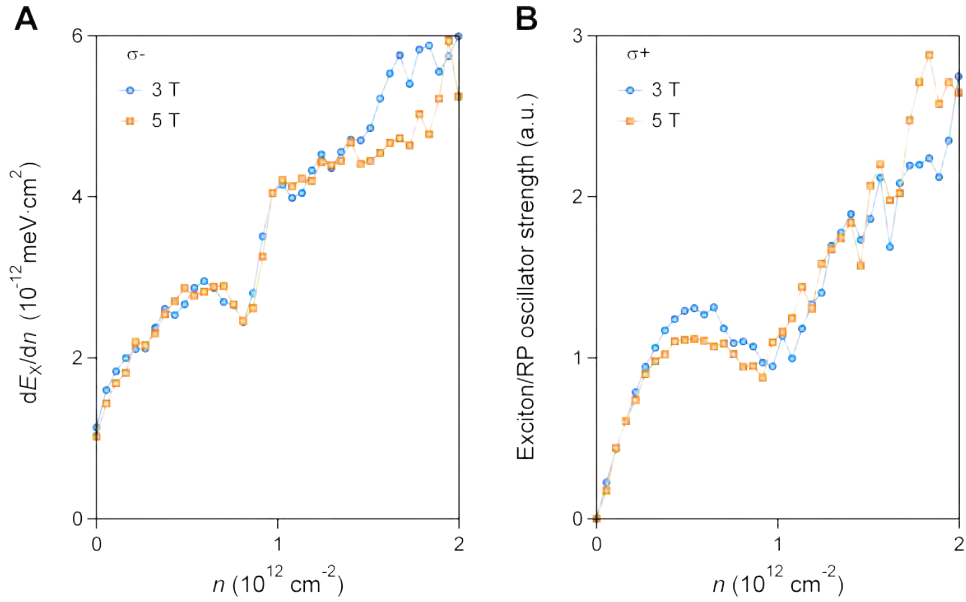

**Fig. S13. Anomalies near $n_*$.** (**A**) Derivatives of the exciton/RP resonance energy with respect to electron density at 3T and 5T for left circularly polarized light. (**B**) Oscillator strength of the exciton/RP resonance at 3T and 5T for right circularly polarized light. Reflectance contrast spectra measurements were performed under a light power of 0.7 nW at a base lattice temperature of 16 mK.

Spin susceptibility as a function of electron density at different temperatures

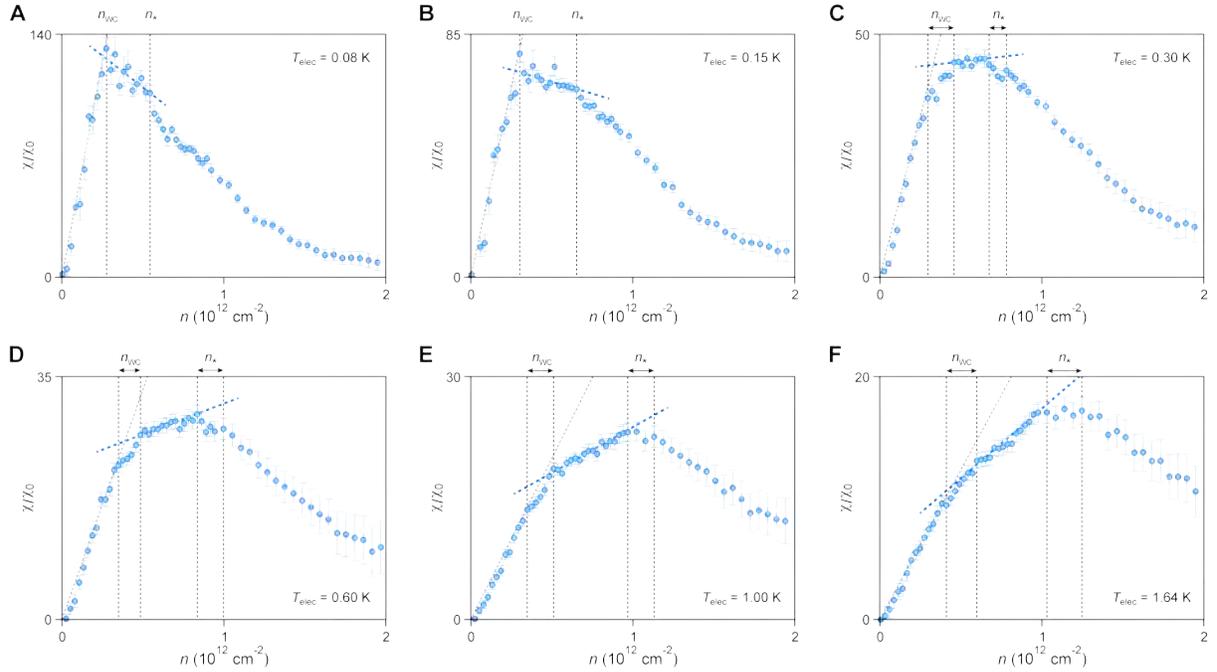

**Fig. S14. Reduced spin susceptibility, the same data as shown in Fig. 3D of the main text.** (**A-F**) The Curie susceptibilities in the Wigner crystal regime are indicated by gray dashed lines. The spin susceptibility between $n_{\mathrm{WC}}$ and $n_*$ aligns well with the colored dashed lines, representing intermediate density ranges that follow the lever rule. However, notable deviations from this linear behavior are observed in the vicinity of $n_{\mathrm{WC}}$ and $n_*$, signifying interface effects between the crystal and liquid regions. The boundaries of these density ranges are delineated by black vertical dashed lines, with these regions further emphasized by black arrows. The density ranges are quantified by the error bars in Fig. 4 of the main text. At low temperatures (A, B), the deviations from the lever rule in the intermediate density ranges become less discernible as the width of the coexistence region diminishes due to the Pomeranchuk effect.

Estimation of electron temperature for the reflectance measurements

In Fig. 3, A to C of the main text, the reflectance contrast spectra were obtained by illuminating the sample with a light power of 70 pW at the base lattice temperature. We use exciton/RP Zeeman energy splitting ($\Delta E_{RP,\pm} = E_{RP,\sigma+} - E_{RP,\sigma-}$) to extract the electron magnetization and estimate the electron temperature from a fit using the Brillouin function.

$\Delta E_{RP,\pm} - \Delta E^0_{RP,\pm}$ ($\Delta E^0_{RP,\pm}$ is bare exciton Zeeman splitting at zero doping) is obtained from the reflectance spectra and we plot it as a function magnetic field at the electron density of $0.49 \times 10^{12}$ cm$^{-2}$ (Fig. S15). We fit the data using the Brillouin function with g-factor of 4.3, which is obtained from the $\widetilde{M}(H)$ curves in the low electron density regime (Fig. 2C in the main text). Estimated electron temperature is 400 mK. We thus mark $n_{WC}(T)$ and $n_*(T)$, determined by reflectance contrast spectra, at 400 mK in (n, T) phase diagram in Fig. 4 in the main text.

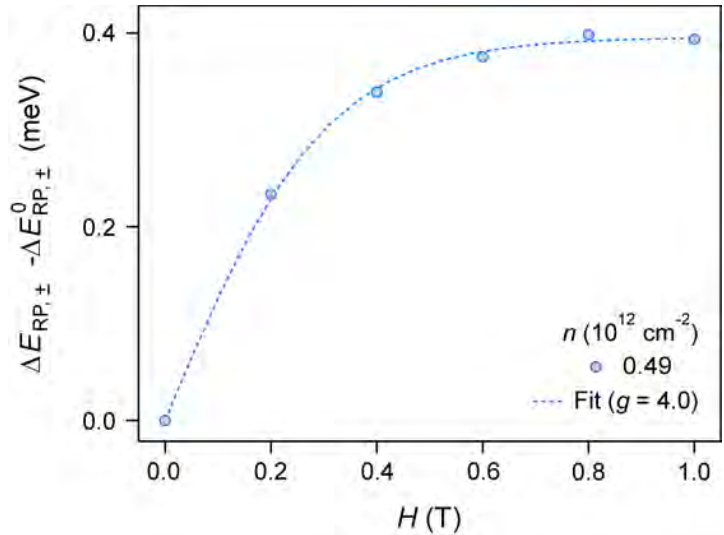

**Fig. S15. Estimation of electron temperature from RP energy splitting.** We plot $\Delta E_{RP,\pm} - \Delta E^0_{RP,\pm}$ as a function of magnetic field for an electron density of 0.49×10$^{12}$ cm$^{-2}$. By using Brillouin function fit, we estimate the electron temperature for the reflectance measurements.

Temperature evolution of the crystal-liquid coexistence

We calculate the temperature evolution of the crystal-liquid coexistence phase from a Maxwell construction. Macroscopic phase separation is forbidden in this system – either for a system with pure Coulomb interactions or with dipolar interactions owing to the presence of a gate electrode. However, if the distance from 2DEG to the gate electrode is sufficiently large (see Eq. 5 in ref (*1*)), the size of the coexisting crystal/liquid regions is big enough such that surface tension between the regions can be neglected. In this case, a Maxwell construction still determines the extent of the coexistence phase, although not its meso/nano-scale structure. The assumption of large coexisting crystal and liquid regions is consistent with the approximate lever-rule behavior of spin susceptibility reported in the main text.

Consider coexistence between a Wigner crystal with density $n_{WC}$ and and liquid with density $n_L$. The densities of individual phases are related to the average density according to $n = (1-x)n_{WC} + xn_L$, where $1-x$ is the areal fraction of the system that is crystal and $x$ is the fraction that is liquid. In our system, charge neutrality is maintained by the gate electrode. We assume a large compressibility of the charges in the gate, implying the associated contribution to the capacitance can be neglected with respect to geometric capacitance of the system. Thus the gate is modeled by purely via its geometric capacitance. The free-energy density of the mixture (ignoring surface tension between regions) is given by

$$f = (1-x)\left[f_{WC}(n_{WC}) + \frac{e^2 n_{WC}^2}{2C}\right] + x\left[f_L(n_L) + \frac{e^2 n_L^2}{2C}\right] \tag{1}$$

where $f_{WC}$ and $f_L$ are the respective free-energy densities of the crystal and liquid phases. The second term in each of the square brackets is the capacitor energy, where the capacitance per unit area is $C = \varepsilon/(4\pi d)$, with $d$ the distance to the gate and $\varepsilon$ the dielectric constant of the environment. The densities $n_{WC}$ and $n_L$ are determined by minimizing the free-energy (1) with respect to $n_{WC}$ and $n_L$. This yields

$$\mu_{WC} + \frac{n_{WC}}{C} = \mu_L + \frac{n_L}{C} = \frac{f_L(n_L) - f_{WC}(n_{WC})}{n_L - n_{WC}} \tag{2}$$

where the chemical potential is $\mu = \partial f/\partial n$.

To study the evolution of phase coexistence at low temperatures we separate the free energies of the two phases into a $T = 0$ part, $\epsilon_{WC/L}(n)$, and a finite temperature correction. For the $T = 0$ contribution we use the parametrized energies from quantum Monte Carlo simulation result (*35*). For temperatures larger than the exchange interaction energy between localized Wigner crystal electrons, the finite temperature contribution to the free-energy in the crystal phase is due to the large spin entropy $S_{WC} \approx n \ln 2$. We thus approximate the free-energy of the crystal phase by $f_{WC}(n) = \epsilon_{WC}(n) - nT \ln 2$. For a Fermi liquid, on the other hand, the entropy is expected to be suppressed $S_{FL} \propto T/E_F^*$, with $E_F^*$ the (renormalized) Fermi energy. Thus, for $T \ll E_F^*$ it is reasonable to approximate the liquid state free-energy just by the zero-temperature energy, $f_L(n) \approx \epsilon_L(n)$. With these approximations we then numerically solve Eqs. (2).

The results of the calculation are summarized in Fig. S16. The magnitude of the Pomeranchuk effect, that is, the slope of the phase boundary with temperature, is comparable to what we observe in experiment (Fig. S16, left panel). As can be seen from the right panel of Fig. S16, we also find the width of the coexistence region increases with increasing temperature, consistent with the experiment. Thus, while the density range over which coexistence occurs is underestimated as compared to the experiment, our calculations suggest the qualitative behavior of the intermediate phase observed in experiment is well captured by a mixed-phase description.

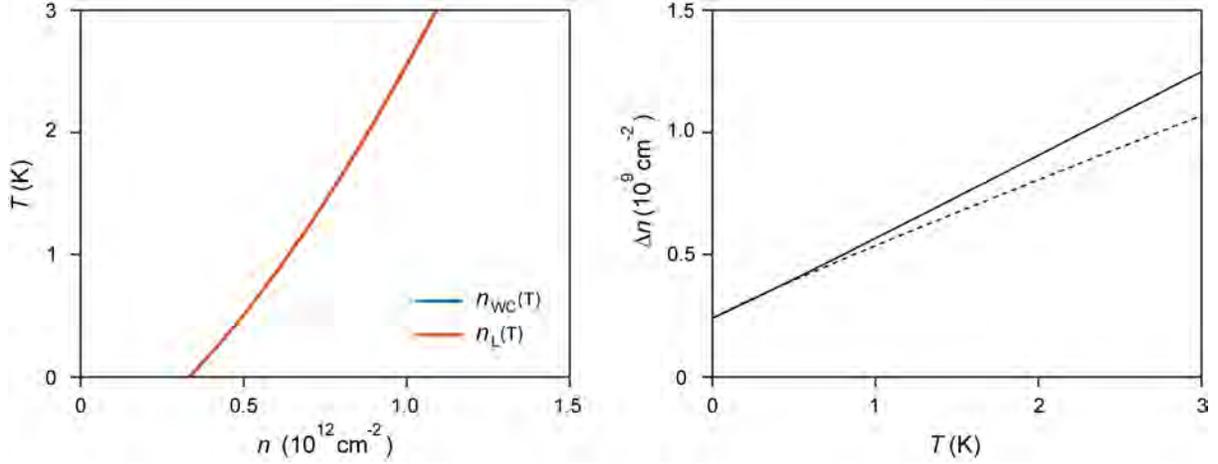

**Fig. S16.** Left, Densities $n_{WC/L}(T)$ delimiting the two-phase coexistence region as obtained from a Maxwell construction. The rightward slope of the curves is consistent with enhanced spin entropy of the crystal phase (Pomeranchuk effect, see text). The width of the coexistence region is not visible on the scale of this plot. Right: Zoom in on the coexistence region. The dashed line shows the analytic approximation to the width of the coexistence region described in the text. Parameters used in the Fig. S16 are $m_e^* = 0.7 m_0$, $\varepsilon = 4.6$, and $d = 15$ nm. The effect of gate electrode is modelled by its geometric capacitance (see text).

To get an analytical handle on the evolution of the coexistence region, it is instructive to make a quadratic approximation to the free-energy in the vicinity of the density $n_c$ ($= (n_{WC} + n_L)/2$) where the zero-temperature energy curves of the two phases cross. We write

$$f_{WC}(n) = f_0 + \mu_{WC}(n - n_c) + \tfrac{1}{2}\alpha_{WC}(n - n_c)^2 - nT \ln 2 + \frac{e^2}{2C}n^2 \tag{3}$$

$$f_L(n) = f_0 + \mu_L(n - n_c) + \tfrac{1}{2}\alpha_L(n - n_c)^2 + \frac{e^2}{2C}n^2 \tag{4}$$

where $\alpha_{WC/L} = d^2\epsilon_{WC/L}(n)/dn^2$ evaluated at $n = n_c$ and $f_0 = f_{WC}(n_c, T = 0) = f_L(n_c)$. Writing the mixed-phase free energy as in Eq. (1) and minimizing with respect to the $n_{WC}$ and $n_L$ yields

$$n_{WC} = n_c - \frac{\delta\mu}{\delta\alpha} + \sqrt{\frac{\alpha_L + e^2/C}{\alpha_{WC} + e^2/C}} \frac{\sqrt{\delta\mu^2 + 2\Delta\delta\alpha}}{\delta\alpha} \tag{5}$$

$$n_L = n_c - \frac{\delta\mu}{\delta\alpha} + \sqrt{\frac{\alpha_{WC} + e^2/C}{\alpha_L + e^2/C}} \frac{\sqrt{\delta\mu^2 + 2\Delta\delta\alpha}}{\delta\alpha} \tag{6}$$

where $\delta\mu = \mu_{WC} - T\ln 2 - \mu_L > 0$ and $\delta\alpha = \alpha_{WC} - \alpha_L > 0$. The former inequality is required for the liquid phase to be favorable at higher densities, while the latter is obtained from QMC results (*35*). We have also defined $\Delta = n_c T \ln 2$. From these results we make a low-$T$ expansion of the width of the coexistence region:

$$\delta n = n_L - n_{WC} \approx \delta n(T=0) + \frac{[n_c(\alpha_{WC}-\alpha_L)-(\mu_{WC}-\mu_L)]\ln 2}{\sqrt{\tilde{\alpha}_{WC}\tilde{\alpha}_L}(\mu_{WC}-\mu_L)} T \qquad (7)$$

Where $\tilde{\alpha}_{WC} = \alpha_{WC} + e^2/C$ and $\tilde{\alpha}_L = \alpha_L + e^2/C$. From the QMC parameterizations we find $n_c(\alpha_{WC} - \alpha_L) - (\mu_{WC} - \mu_L) > 0$, so that the coexistence regions indeed widens upon increasing $T$. We note that the difference of compressibilities between the crystal and liquid phases is crucial to reproduce the experimentally observed dependence of the coexistence region's width on temperature.

Spin polarization of the 2DEG from quantum Monte Carlo

To determine the theoretical evolution of the electronic spin polarization in an applied magnetic field $H$ (Fig. 2A), we have utilized existing QMC parametrizations of the ground state energy of the 2DEG as a function of density $(n)$ and polarization $\zeta$. Ignoring orbital effects, which are small for the densities and magnetic field strengths under consideration, we write the energy per particle of the 2DEG in a magnetic field as $E(n, \zeta, H) = E(n, \zeta, H = 0) - \frac{1}{2}g\mu_B\zeta H$, and obtain the spin polarization by minimizing the energy: $dE(n, \zeta, H)/d\zeta = 0$, which determines $\zeta(H)$. The function $E(n, \zeta, H = 0)$ was previously parametrized by fitting to QMC simulations (*35*).

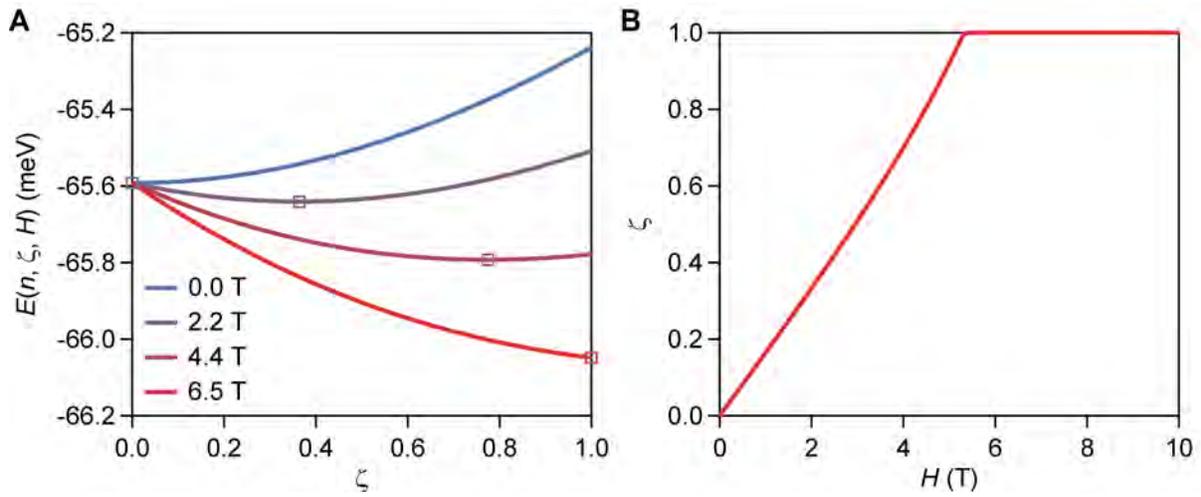

**Fig. S17. Spin polarization of the 2DEG from quantum Monte Carlo.** (**A**) Energy per particle of the 2DEG in a magnetic field as a function of spin polarization at various fields and electron density $1.8 \times 10^{12}$ cm$^{-2}$. Square markers indicate the spin polarization at minimum energy for each field. (**B**) Spin polarization of the 2DEG, determined by minimizing energy as in (A), as a function of magnetic field at electron density $1.8 \times 10^{12}$ cm$^{-2}$.

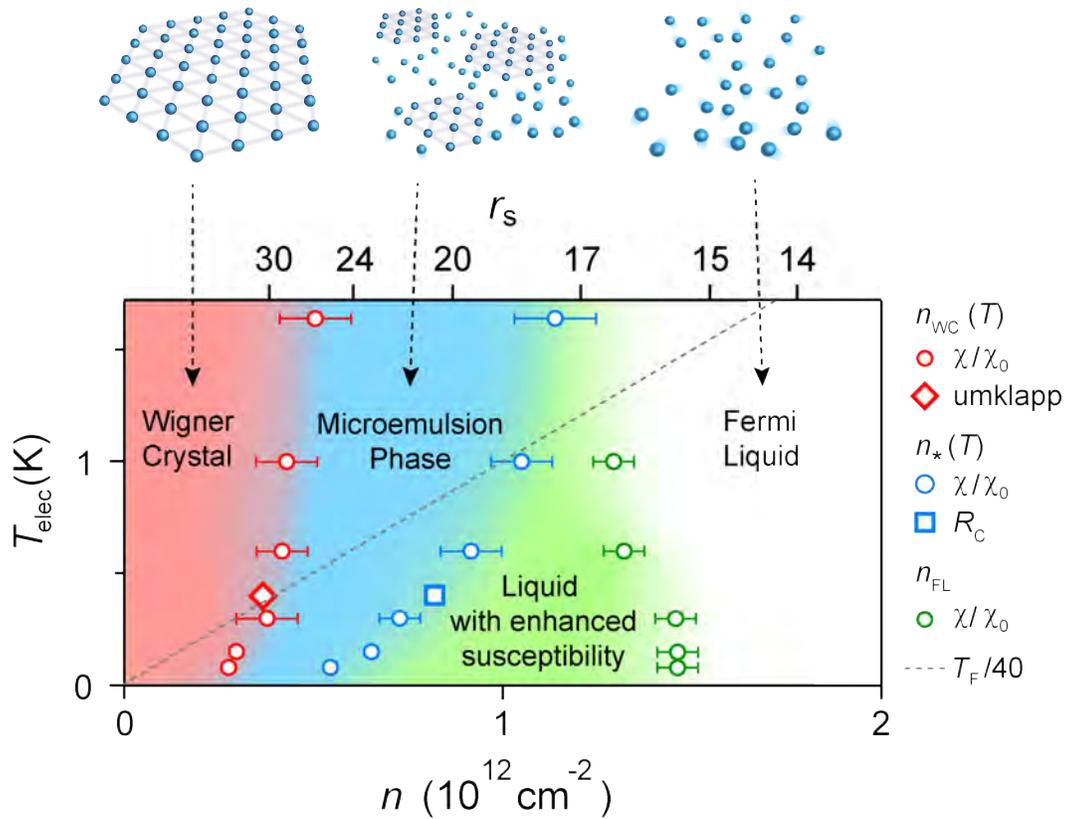

**Fig. S18: Phase diagram as a function of electron density and temperature.** The gray dashed line indicates the density-dependent Fermi temperature of the non-interacting 2DEG, which is divided by a factor of 40 to align with the temperature scale of the phase diagram.